\documentclass[twocolumn,superscriptaddress,prl,showpacs,amsmath,amssymb,letterpaper]{revtex4-1}
\usepackage{graphicx,color}
\usepackage{amssymb}
\usepackage{amsmath}
\usepackage{latexsym}
\usepackage{amsxtra}
\usepackage{amscd}
\usepackage{dcolumn}
\usepackage{bm}
\usepackage{threeparttable}
\usepackage[breaklinks]{hyperref}
\usepackage{url}
\usepackage{breakurl}
\graphicspath{{./}}
\let\oldsqrt\sqrt
\def\sqrt{\mathpalette\DHLhksqrt}
\def\DHLhksqrt#1#2{\setbox0=\hbox{$#1\oldsqrt{#2\,}$}\dimen0=\ht0
\advance\dimen0-0.2\ht0
\setbox2=\hbox{\vrule height\ht0 depth -\dimen0}%
{\box0\lower0.4pt\box2}}

\newcommand{\nuc}[2]{$^{#1}$#2}

\newcommand{\bev}{$B(E2)$ value}
\newcommand{\bevs}{$B(E2)$ values}

\newcommand{\bevups}{$B(E2;\;0^+ \rightarrow 2^+)$ values}
\newcommand{\bevupn}[1]{$B(E2;\;0^+_{#1} \rightarrow 2^+_{#1})$ value}
\newcommand{\bevupns}[1]{$B(E2;\;0^+_{#1} \rightarrow 2^+_{#1})$ values}
\newcommand{\beup}{$B(E2;\;0^+ \rightarrow 2^+)$}

\begin{document}
\title{Shape changes in the mirror nuclei \nuc{70}{Kr} and \nuc{70}{Se}}
\author{K.~Wimmer}
\email{Corresponding author: k.wimmer@csic.es}
\affiliation{Instituto de Estructura de la Materia, CSIC, E-28006 Madrid, Spain}
\affiliation{Department of Physics, The University of Tokyo, Hongo, Bunkyo-ku, Tokyo 113-0033, Japan}
\affiliation{RIKEN Nishina Center, 2-1 Hirosawa, Wako, Saitama 351-0198, Japan}
\author{W.~Korten} 
\affiliation{IRFU, CEA, Universit\'{e} Paris-Saclay, F-91191 Gif-sur-Yvette, France}
\author{P.~Doornenbal} 
\affiliation{RIKEN Nishina Center, 2-1 Hirosawa, Wako, Saitama 351-0198, Japan}
\author{T.~Arici} 
\affiliation{GSI Helmholtzzentrum f\"{u}r Schwerionenforschung, D-64291 Darmstadt, Germany}
\affiliation{Justus-Liebig-Universit\"{a}t Giessen, D-35392 Giessen, Germany}
\author{P.~Aguilera} 
\affiliation{Comisi\'{o}n Chilena de Energ\'{i}a Nuclear, Casilla 188-D, Santiago, Chile}
\author{A.~Algora}
\affiliation{Instituto de Fisica Corpuscular, CSIC-Universidad de Valencia, E-46071 Valencia, Spain}
\affiliation{Institute of Nuclear Research of the Hungarian Academy of Sciences, Debrecen H-4026, Hungary}
\author{T.~Ando} 
\affiliation{Department of Physics, The University of Tokyo, Hongo, Bunkyo-ku, Tokyo 113-0033, Japan}
\author{H.~Baba} 
\affiliation{RIKEN Nishina Center, 2-1 Hirosawa, Wako, Saitama 351-0198, Japan}
\author{B.~Blank} 
\affiliation{CENBG, CNRS/IN2P3, Universit\'{e} de Bordeaux, F-33175 Gradignan, France}
\author{A.~Boso} 
\affiliation{Istituto Nazionale di Fisica Nucleare, Sezione di Padova, I-35131 Padova, Italy}
\author{S.~Chen} 
\affiliation{RIKEN Nishina Center, 2-1 Hirosawa, Wako, Saitama 351-0198, Japan}
\author{A.~Corsi} 
\affiliation{IRFU, CEA, Universit\'{e} Paris-Saclay, F-91191 Gif-sur-Yvette, France}
\author{P.~Davies} 
\affiliation{Department of Physics, University of York, YO10 5DD York, United Kingdom}
\author{G.~de~Angelis} 
\affiliation{Istituto Nazionale di Fisica Nucleare, Laboratori Nazionali di Legnaro, I-35020 Legnaro, Italy}
\author{G.~de~France} 
\affiliation{GANIL, CEA/DSM-CNRS/IN2P3, F-14076 Caen Cedex 05, France}
\author{J.-P.~Delaroche} 
\affiliation{CEA, DAM, DIF, F-91297 Arpajon, France}
\author{D.~T.~Doherty}
\affiliation{IRFU, CEA, Universit\'{e} Paris-Saclay, F-91191 Gif-sur-Yvette, France}
\author{J.~Gerl} 
\affiliation{GSI Helmholtzzentrum f\"{u}r Schwerionenforschung, D-64291 Darmstadt, Germany}
\author{R.~Gernh\"{a}user} 
\affiliation{Physik Department, Technische Universit\"{a}t M\"{u}nchen, D-85748 Garching, Germany}
\author{M.~Girod}
\affiliation{CEA, DAM, DIF, F-91297 Arpajon, France}
\author{D.~Jenkins} 
\affiliation{Department of Physics, University of York, YO10 5DD York, United Kingdom}
\author{S.~Koyama} 
\affiliation{Department of Physics, The University of Tokyo, Hongo, Bunkyo-ku, Tokyo 113-0033, Japan}
\author{T.~Motobayashi} 
\affiliation{RIKEN Nishina Center, 2-1 Hirosawa, Wako, Saitama 351-0198, Japan}
\author{S.~Nagamine} 
\affiliation{Department of Physics, The University of Tokyo, Hongo, Bunkyo-ku, Tokyo 113-0033, Japan}
\author{M.~Niikura} 
\affiliation{Department of Physics, The University of Tokyo, Hongo, Bunkyo-ku, Tokyo 113-0033, Japan}
\author{A.~Obertelli} 
\altaffiliation[Present address: ]{Institut f\"ur Kernphysik Technische Universit\"at Darmstadt Germany}
\affiliation{IRFU, CEA, Universit\'{e} Paris-Saclay, F-91191 Gif-sur-Yvette, France}
\author{J.~Libert}
\affiliation{CEA, DAM, DIF, F-91297 Arpajon, France}
\author{D.~Lubos} 
\affiliation{Physik Department, Technische Universit\"{a}t M\"{u}nchen, D-85748 Garching, Germany}
\author{T.~R.~Rodr{\'i}guez}
\affiliation{Departamento de F\'isica Te\'orica and Centro de Investigaci\'on Avanzada en F\'isica Fundamental, Universidad Aut{\'o}noma de Madrid, E-28049 Madrid, Spain}
\author{B.~Rubio} 
\affiliation{Instituto de Fisica Corpuscular, CSIC-Universidad de Valencia, E-46071 Valencia, Spain}
\author{E.~Sahin} 
\affiliation{Department of Physics, University of Oslo, PO Box 1048 Blindern, N-0316 Oslo, Norway}
\author{T.~Y.~Saito} 
\affiliation{Department of Physics, The University of Tokyo, Hongo, Bunkyo-ku, Tokyo 113-0033, Japan}
\author{H.~Sakurai} 
\affiliation{Department of Physics, The University of Tokyo, Hongo, Bunkyo-ku, Tokyo 113-0033, Japan}
\affiliation{RIKEN Nishina Center, 2-1 Hirosawa, Wako, Saitama 351-0198, Japan}
\author{L.~Sinclair} 
\affiliation{Department of Physics, University of York, YO10 5DD York, United Kingdom}
\author{D.~Steppenbeck} 
\affiliation{RIKEN Nishina Center, 2-1 Hirosawa, Wako, Saitama 351-0198, Japan}
\author{R.~Taniuchi} 
\affiliation{Department of Physics, The University of Tokyo, Hongo, Bunkyo-ku, Tokyo 113-0033, Japan}
\author{R.~Wadsworth} 
\affiliation{Department of Physics, University of York, YO10 5DD York, United Kingdom}
\author{M.~Zielinska}
\affiliation{IRFU, CEA, Universit\'{e} Paris-Saclay, F-91191 Gif-sur-Yvette, France}

\begin{abstract}
  We studied the proton-rich $T_z=-1$ nucleus \nuc{70}{Kr} through inelastic scattering at intermediate energies in order to extract the reduced transition probability, $B(E2;\;0^+ \rightarrow 2^+)$. Comparison with the other members of the $A=70$ isospin triplet, \nuc{70}{Br} and \nuc{70}{Se}, studied in the same experiment, shows a $3\sigma$ deviation from the expected linearity of the electromagnetic matrix elements as a function of $T_z$. 
  At present, no established nuclear structure theory can describe this observed deviation quantitatively. This is the first violation of isospin symmetry at this level observed in the transition matrix elements. 
A heuristic approach may explain the anomaly by a shape  change between the mirror nuclei \nuc{70}{Kr} and \nuc{70}{Se} contrary to the model predictions.
\end{abstract}
\date{\today}
\maketitle
The strong interaction is independent of the electric charge of a particle, its Hamiltonian commutes with the isobaric spin operator $T$.  Within this isobaric spin symmetry, the proton ($T_z=-1/2$) and the neutron ($T_z=+1/2$) are two representations of a particle, the nucleon~\cite{heisenberg32}. 
Electromagnetic effects violate isobaric spin symmetry and the light quark mass difference ($m_\text{u}\neq m_\text{d}$) results in a larger neutron mass than the mass of a proton~\cite{miller06} making the neutron unstable. 
The relative mass difference between neutrons and protons is only 0.0013, suggesting that the symmetry breaking related to the strong interaction is rather small and the observable effects are dominated by the electromagnetic interaction.
Precise measurements of the $nn$, $pp$, and $np$ scattering length, for example, and careful correction of all electro-magnetic effects nevertheless demonstrated that proton-proton ($T_z=-1$), neutron-proton ($T_z=0$), or neutron-neutron ($T_z=+1$) interactions are different~\cite{miller06}.

In atomic nuclei, the charge independence of the nuclear interaction implies: (i) exactly degenerate energies of isobaric multiplets~\cite{wigner37}, (ii) pure isospin quantum numbers and no isospin mixing in nuclear states, and (iii) identical wave functions for the members of an isobaric multiplet. 
For charge dependent two-body interactions the masses of isobaric nuclei depend on the isospin projection $T_z$. The isobaric multiplet mass equation (IMME) relates the mass-excess of isobars as a quadratic function of $T_z$. Deviations from the IMME indicate isospin mixing, isospin symmetry breaking or the presence of three-body forces.

Isospin symmetry is typically studied through Mirror Energy Differences (MED) to test the charge symmetry of the nuclear interaction and Triplet Energy Differences (TED) to test charge independence of the nuclear interaction. Very recently, isospin symmetry breaking has been observed in the $A=73$ mirror pair, where the ground state spins of \nuc{73}{Sr} and \nuc{73}{Br} differ~\cite{hoff20}. The excitation energy difference of the first two states in \nuc{73}{Br} is, however, only 27~keV, so that the absolute scale of this violation is very small and comparable to other cases~\cite{bentley07}. The analysis of mirror energy differences of excited states shows that isospin non-conserving interactions are required in addition to the Coulomb force to reproduce the observation with shell model calculations~\cite{zuker02}. The origin of the additional, phenomenological terms in the interaction is not yet understood.
Excitation energies alone, however, do not reveal the isospin purity of states and do not probe the mirror symmetry of the wave functions.

An alternative and more rigorous way to test isospin symmetry are electromagnetic matrix elements. In contrast to the excitation energies, the matrix elements also probe the properties (ii) and (iii) of the charge independence of the nuclear interaction. In such a case the isospin dependence of the proton matrix element for a $T=1$ triplet is given by a simple linear relation~\cite{bernstein79}
\begin{equation}
  M_\text{p}(T_z) = \frac{1}{2}\left( M_0 - M_1 T_z\right)
\label{eq:linmat}
\end{equation}
with the isoscalar ($M_0$) and isovector ($M_1$) matrix elements.
Experimentally, this linearity can be tested through measurements of the reduced electromagnetic transition probability
\begin{equation}
  B(E\lambda;\; J_\text{i} \rightarrow J_\text{f} ) = \frac{|M_\text{p}(T_z)|^2}{2J_\text{i} + 1}
\label{eq:be2mat}
\end{equation}
of the decay of the first $T,J^\pi = 1,2^+$ to $1,0^+$ states in the triplet. This has been studied for $T=1$ triplets with $22\leq A \leq 50$~\cite{pradosestevez07,boso19,giles19} and no deviation from the expected linear trend was detected within the experimental uncertainties.
Isospin mixing of $T=0$ and $T=1$ states in the odd-odd system could potentially disturb the linearity, but different systematic uncertainties from different experiments make it difficult to draw conclusions~\cite{pradosestevez07}.

In this letter, we present the first case where the electromagnetic matrix elements significantly deviate from the linear trend of Eq.~\ref{eq:linmat}. The $A=70$ nuclei have been chosen for this because previous experimental investigations of the Coulomb energy differences of \nuc{70}{Se} and \nuc{70}{Br} found an anomalous behavior~\cite{deangelis01}, which could be interpreted as a shape change between the isobars~\cite{narasingh07}. Spectroscopy of \nuc{70}{Kr} at $T_z=-1$ was only achieved recently~\cite{debenham16,wimmer18}, but did not allow to probe the nuclear shape and test for a proposed shape change in the mirror pair \nuc{70}{Se} and \nuc{70}{Kr}~\cite{petrovici15}. Electromagnetic transition matrix elements, on the other hand, also allow for the determination of the shape or deformation of a nucleus. In a rotational model, the $B(E2)$ value is related to the magnitude of the (intrinsic) quadrupole moment $Q_0$, and the absolute value of the deformation $\beta_2$~\cite{bohr75}.

The experiment was performed at the Radioactive Isotope Beam Factory, operated by RIKEN Nishina Center and CNS, University of Tokyo. Nuclei along the $N=Z$ line were produced by projectile fragmentation of an intense \nuc{78}{Kr} primary beam at an energy of 345~$A$MeV. The reaction products were selected and identified in the \mbox{BigRIPS} separator~\cite{kubo12} using the $B\rho-\Delta E-TOF$ method. The average intensity of the \nuc{70}{Kr} beam was 15~pps with a total fraction of 0.9\% in the secondary beam.
At the final focus of the \mbox{BigRIPS} separator the beam impinged on 926(2)~mg/cm$^2$ thick Au and 703(7)~mg/cm$^2$ thick Be targets. The targets were surrounded by the DALI2 detector array~\cite{takeuchi14}, consisting of 186 individual NaI(Tl) crystals. Reaction products were identified in the ZeroDegree spectrometer~\cite{kubo12} using the same technique as for BigRIPS.
Further details of the experiment can be found in \cite{wimmer18,wimmer20}.

\begin{figure}[ht]
  \centering
\includegraphics[width=\columnwidth]{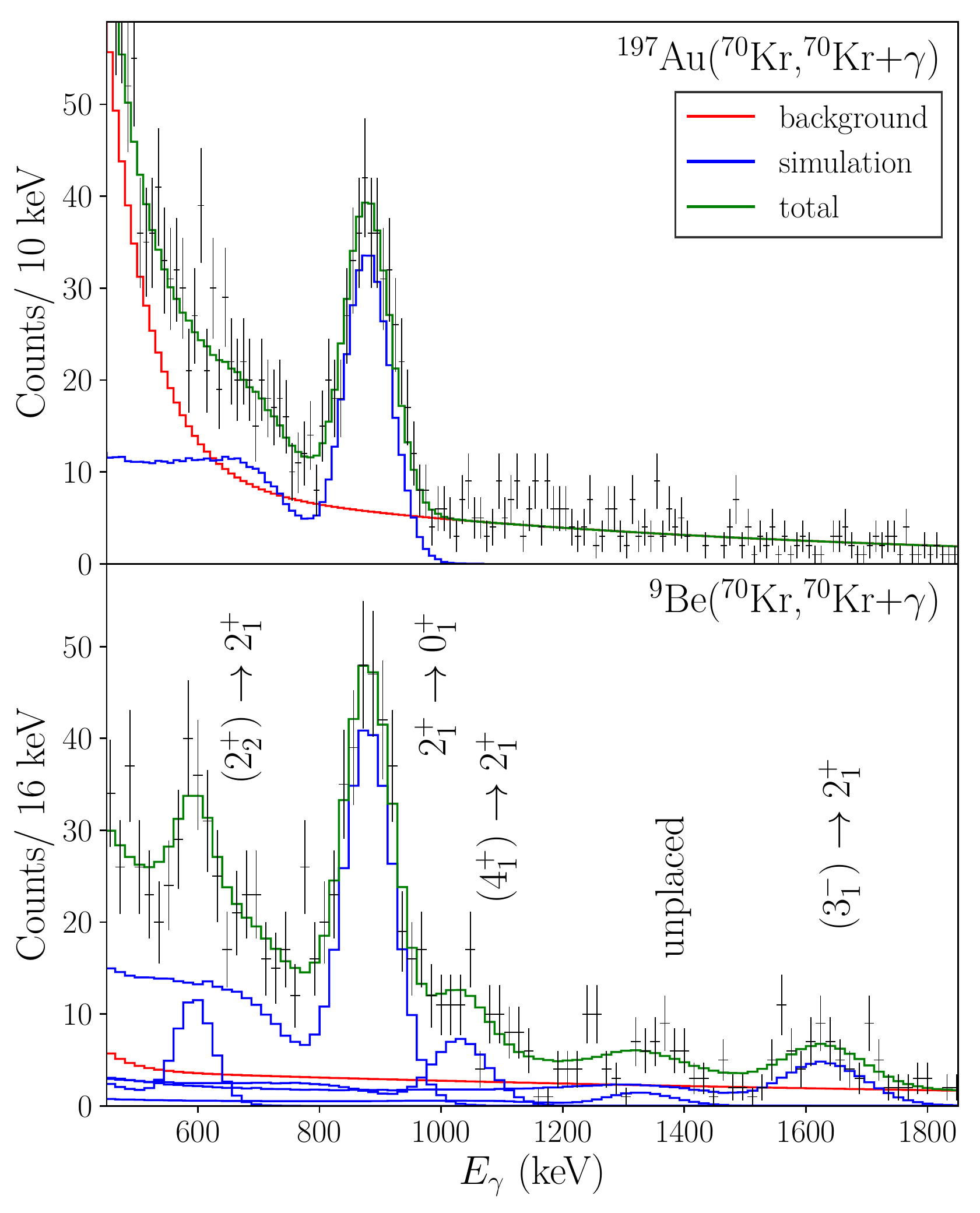}
\caption{Doppler-corrected $\gamma$-ray energy spectrum for the inelastic scattering of \nuc{70}{Kr} on a \nuc{197}{Au} (top) and \nuc{9}{Be} (bottom) targets. The Doppler correction assumes $\gamma$-ray emission at the velocity behind the target. The data are fitted with simulated response functions for the transition at 884~keV and a continuous background (red). For the Au target data, only forward DALI2 crystals ($\theta_\text{lab}<75^\circ$) are shown to reduce background from atomic processes. For the Be target data see Ref.~\cite{wimmer18} for further details.}
\label{fig:auspectrum}
\end{figure}

The Doppler-corrected $\gamma$-ray energy spectra for \nuc{70}{Kr} impinging on the Au and Be targets are shown in Fig.~\ref{fig:auspectrum}.
In both spectra the decay of the 885~keV excited $2^+$ state to the ground state is observed.
The $\gamma$-ray yield has been determined by fitting a simulated response function and a continuous double exponential background to the data. The simulated response function assumes a level lifetime of 2.8~ps, consistent with the results extracted from the Coulomb excitation cross section (see below).
For the scattering with the Au target, only the data from the forward DALI2 crystals have been taken into account to reduce the low-energy background from atomic processes. The angular distribution for the $E2$ transition was included in the simulation.
In order to extract the exclusive cross section for the excitation of the $2^+_1$ state, indirect feeding from higher-lying states has to be subtracted. For the inelastic scattering off the Be target, the yield for states above the $2^+_1$ state was subtracted to correct for indirect feeding as described in Ref.~\cite{wimmer18}. For the data taken with the Au target, the de-excitation from the $2^+_2$ state at 1478 keV to the $2^+_1$ state is not observed.
The cross section for the excitation of the $2^+_1$ state in \nuc{70}{Kr} amounts to 349(36)~mb (see Table~\ref{tab:cs}).
Adding a transition at 594~keV to the fit of the spectrum shown in the top panel of Fig.~\ref{fig:auspectrum} results in an upper limit for the cross section $\sigma(2^+_2)$ of 15~mb. States above the $2^+_2$ state are expected to contribute even less. In order to account for them and the uncertainty related to the feeding of the $2^+_2$ state an additional 15~mb has been added to the systematic uncertainty. 
The corrections and previously quoted systematic uncertainty for the cross section are taken into account when the \bevups\ are determined from the measured cross sections.

Within the same spectrometer setting, also the isobars \nuc{70}{Br} ($E(2^+_1) = 934$~keV) and \nuc{70}{Se} ($E(2^+_1) = 954$~keV) were transmitted. For the former, an isomeric $9^+$ state at 2292~keV~\cite{karny04} allows, in principle, to only extract a lower limit for the excitation cross sections of the $2^+_1$ state. However, no transition besides the $2^+_1\rightarrow 0^+_1$ decay has been observed in the scattering off the Au target, indicating a small isomeric ratio in the beam or a small $B(E2)$ value for the states above the isomer.
For \nuc{70}{Se}, statistics are limited because the acceptance of BigRIPS was optimized for the more exotic $N<Z$ nuclei. 
Based on the systematics of the less exotic Kr isotopes, a low-lying excited $0^+$ state might be present in the beam as an isomeric contamination. In the mirror nucleus \nuc{70}{Se} no such state is known. The $0^+_2$ state candidate is located at 2010~keV~\cite{wadsworth80} and is thus short lived. Theoretical calculations predict the $0^+_2$ in \nuc{70}{Kr} at considerably higher energy than the $2^+_1$ state~\cite{delaroche10,rodriguez14} as well, so that its lifetime would be much shorter than the flight time to the BigRIPS focal plane.
In the analysis of nucleon removal reactions from the same experiment no evidence for a low-lying $0^+$ state was found in either of the two nuclei~\cite{wimmer18}.
Due to the low beam intensity, it was not possible to search for an isomeric state in \nuc{70}{Kr} as it was done for \nuc{72}{Kr}~\cite{wimmer20}, where an isomeric ratio of $4(1)$\% was found.
In the following extraction of the \bevs\ it was assumed that the $A=70$ beam particles are in their respective ground state, when reaching the secondary reaction target.

The cross sections measured with both targets for all three beams are listed in Table~\ref{tab:cs}.
\begin{table}[ht]
  \caption{Measured cross sections and deduced nuclear deformation length $\delta_\text{N}$, proton matrix elements $M(E2)$, and \bevups\ for the $A=70,T=1$ triplet. The total uncertainties are listed together with the individual contributions of statistical, systematic, and theoretical uncertainties.}
\begin{ruledtabular}
  \begin{tabular}{lccc}
    & \nuc{70}{Kr} & \nuc{70}{Br} & \nuc{70}{Se}  \\
    $E(2^+)$ (keV) & 885 & 934 & 954 \\ 
    $\sigma(2^+_1)_\text{Be}$ (mb) & 14.5(46)(10) & 14.6(5)(8) & 15.1(53)(28) \\
    $\sigma(2^+_1)_\text{Au}$ (mb) & 349(36)(27)  & 201(11)(20) & 234(70)(43) \\
    \hline
    \hline
    $\delta_\text{N}$ (fm) & 1.10(19) & 1.09(3) & 1.11(22)\\
    $\Delta^\text{stat}\delta_\text{N}$ (fm) &0.19 & 0.02 & 0.20\\
    $\Delta^\text{syst}\delta_\text{N}$ (fm) &0.04 & 0.03 & 0.10\\
    \hline
    $M(E2)$ (efm$^2$)      & 52.2(43) & 38.1(31) & 40.7(82)\\
    $\Delta^\text{stat}M(E2)$ (efm$^2$)      & 2.8 & 1.2 & 6.8\\
    $\Delta^\text{syst}M(E2)$ (efm$^2$)      & 2.1 & 2.2 & 4.1\\
    $\Delta^\text{theo}M(E2)$ (efm$^2$)      & 2.5 & 1.8 & 2.0\\    
    \hline
    $B(E2)$ (e$^2$fm$^4$)    & 2726(451) & 1454(233) & 1659(659) \\
    $\Delta^\text{stat}B(E2)$ (e$^2$fm$^4$)    & 294 & 91  & 543 \\
    $\Delta^\text{syst}B(E2)$ (e$^2$fm$^4$)    & 224 & 165 & 336 \\
    $\Delta^\text{theo}B(E2)$ (e$^2$fm$^4$)    & 258 & 137 & 164 \\
  \end{tabular}
  \label{tab:cs}
\end{ruledtabular}
\end{table}
Besides the statistical uncertainty resulting from the fitting of the $\gamma$-ray spectrum and the subtraction of feeding in case of the Be target data, a number of systematic uncertainties contribute to the total uncertainty for the cross section. These include the full-energy peak detection efficiency of DALI2 (5\%), target thickness (1\%), ZeroDegree efficiency and transmission (3\% for \nuc{70}{Kr} and \nuc{70}{Br}, 10\% for \nuc{70}{Se}), trigger efficiency (2\%), effects of the $\gamma$-ray angular distribution (2\%), and the unobserved indirect feeding discussed above.

The excitation of the $2^+$ states of interest is caused by both the electro-magnetic and the nuclear interaction between target and projectile. These two contributions interfere and can not be disentangled. The extraction of the nuclear deformation length $\delta_\text{N}$ and the reduced transition probability $B(E2)$ from the measured cross sections requires therefore a consistent reaction model analysis. The procedure is described in~\cite{wimmer20} in detail. Reaction model calculations were performed with a modified version of the distorted wave coupled channels code FRESCO~\cite{thompson88, *moro18} using optical model potentials calculated using the method described in~\cite{furumoto12}. Both the input nuclear deformation length and the \bev\ for the projectile nucleus were adjusted to reproduce simultaneously the measured cross sections for the Be and Au target. The resulting nuclear deformation lengths and $E2$ matrix elements are listed in Table~\ref{tab:cs}. In addition to the statistical and systematic uncertainties discussed above, also uncertainties related to the reaction model and its input parameters are taken into account. The sources consist of the optical model potential (8\%), the treatment of relativistic dynamics (5\%), and uncertainties in the determination of $\delta_\text{N}$ which propagate to the determination of the \bevs.
A detailed discussion of the experimental and theoretical uncertainties is presented in~\cite{wimmer20}. 

\begin{figure}[ht]
  \centering
\includegraphics[width=\columnwidth]{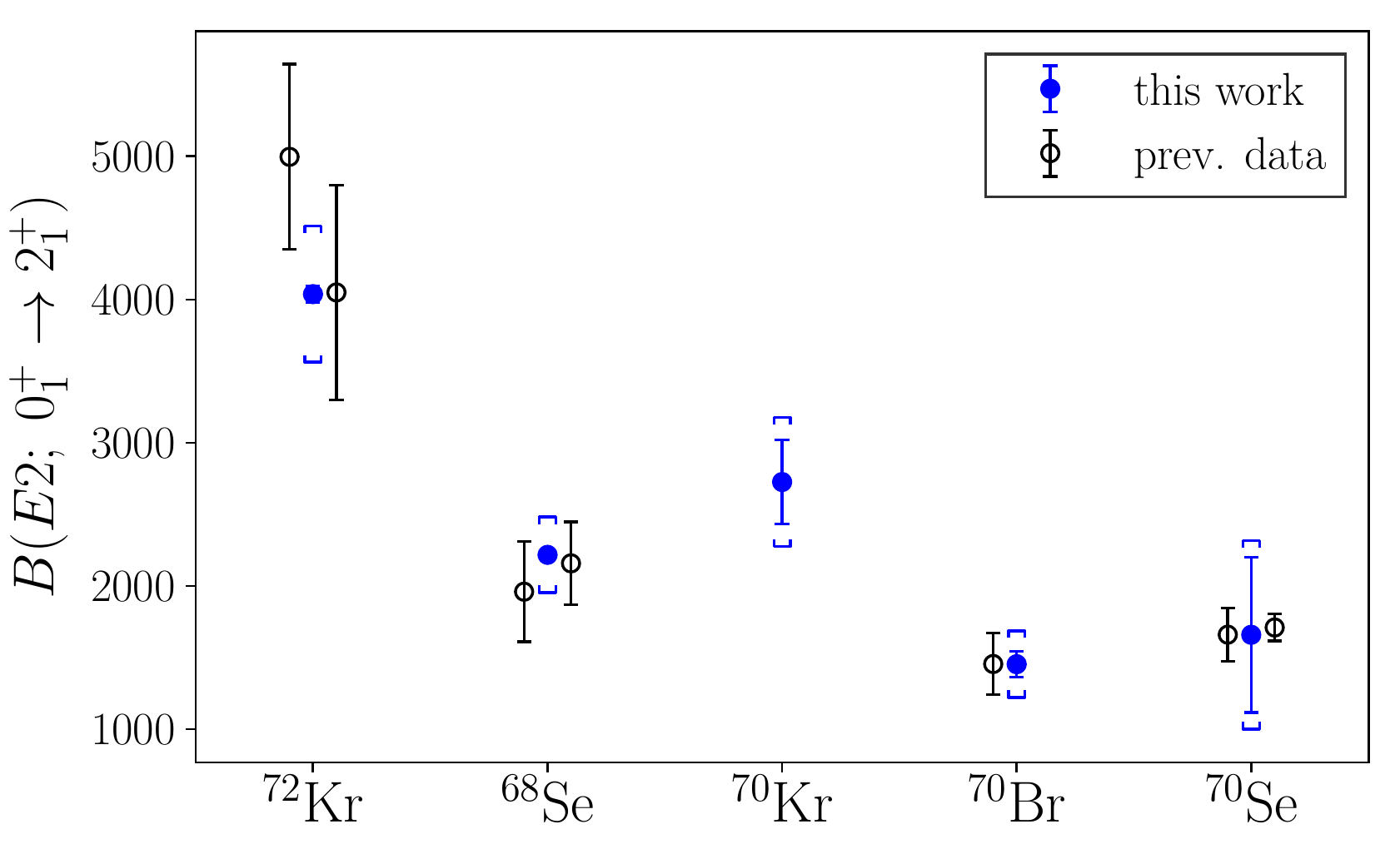}
\caption{Summary of the results for the \bevupns{1} extracted in the present work. The results for \nuc{72}{Kr} were already presented in~\cite{wimmer20}. The error bars indicate statistical uncertainties, while the additional caps show the total uncertainties including statistical, systematical, and uncertainties arising from the reaction theory calculations. For \nuc{72}{Kr} and \nuc{68}{Se} the statistics uncertainty is smaller than the symbol size. Previous experimental results are taken from~\cite{gade05,iwasaki14,obertelli09,nichols14,ljungvall08} and shown by the open symbols.}
\label{fig:bevs}
\end{figure}

In order to validate the analysis procedure, the results for the present experiment are compared to previous measurements of neighboring nuclei using both Coulomb excitation and lifetime measurements. The \bevs\ for the $N=Z$ nuclei \nuc{72}{Kr} and \nuc{68}{Se} as well as the $A=70$ isobars are shown in Fig.~\ref{fig:bevs}. In all four cases the present results agree with the previous measurements. 

\begin{figure}[ht]
  \centering
\includegraphics[width=\columnwidth]{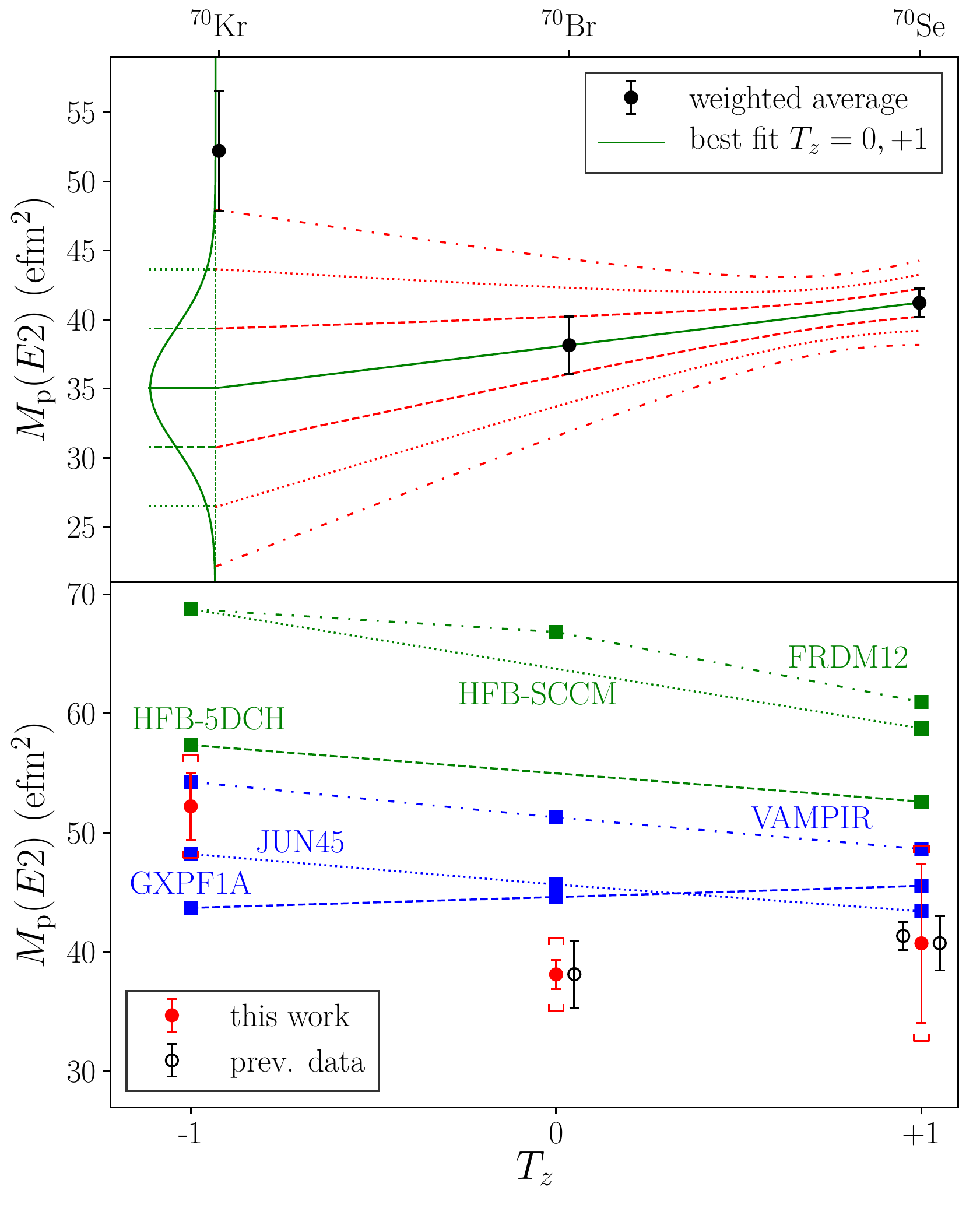}
\caption{$E2$ matrix elements as a function of isospin projection $T_z$ for the $A=70$ nuclei.
  (Top panel) Weighted averages of the present and previous data are shown by the black data points. The linear fit of the data points at $T_z=+1$ and 0 is shown by the solid green line.
  The red dashed, dotted, and dash-dotted lines show the 1, 2, and 3 $\sigma$ confidence intervals of the linear fit.
  The extrapolation of the probability distribution to $T_z=-1$ is shown by the Gaussian curve in green. 
  (Bottom panel) The present results are shown in red full circles and compared to previous data~\cite{nichols14,ljungvall08} (black open circles). The data points are slightly offset on the $T_z$ axis for visualization purposes only. Squares show the results of the theoretical calculations shown in Table~\ref{tab:theory}. } 
\label{fig:mats}
\end{figure}

The matrix elements for the $A=70$ triplet are shown in Fig.~\ref{fig:mats}. 
It can be seen that the value for \nuc{70}{Kr} clearly deviates from the negative trend indicated by the previously known $M_\text{p}$ values for \nuc{70}{Br} and \nuc{70}{Se}. A linear fit for these latter two nuclei with Eq.~\ref{eq:linmat} results in $M_0 = 76(4)$~efm$^2$ and $M_1 = -6(5)$~efm$^2$.
In order to gauge the deviation from this trend, the confidence interval has been determined. 
The weighted average of the previously and the presently determined value for \nuc{70}{Br}~\cite{nichols14} as well as the weighted average of two previous measurements for \nuc{70}{Se}~\cite{nichols14,ljungvall08} were fitted using  linear regression according to Eq.~\ref{eq:linmat} and then extrapolated to $T_z=-1$.
The result of the extrapolation, shown by the green Gaussian curve in the top panel Fig.~\ref{fig:mats} amounts to $M(E2,T_z=-1) = 35.0(43)$~efm$^2$. The experimental value for \nuc{70}{Kr} ($M(E2) = 52.2(43)$~efm$^2$) deviates by more than 3 $\sigma$ from this extrapolation.

In many cases, especially for medium heavy nuclei, the \bev\ for the proton-rich $T_z-1$ nucleus is not known experimentally. Therefore, the isoscalar and isovector matrix elements $M_0$ and $M_1$ were extracted from a fit of Eq.~\ref{eq:linmat} only to the $T_z=0$ and $+1$ members of the triplet~\cite{morse18}. The isovector matrix element $M_1$ was found to be very small. For the cases with $A\geq 50$, where known, the values of $M_1$ are all slightly negative, albeit compatible, within errors, with zero~\cite{morse18}. The work was extended to include the newest available data for $A=78$~\cite{lemasson12,llewellyn20} and a negative isoscalar matrix element was found again. The negative trend for the $T_z=0$ and $+1$ members observed for the present $A=70$ case is thus not unique.

If the $M_p$ data for all three values of $T_z$ are fitted with a curve using simple linear regression, both the matrix elements for \nuc{70}{Br} and \nuc{70}{Kr} deviate by about $2\sigma$ from the curve.
Fitting the matrix elements shown in Fig.~\ref{fig:mats} by a quadratic curve ($M_\text{p} = a+bT_z+cT_z^2$) such as suggested in Ref.~\cite{boso19} to test isospin symmetry results in a $c=8.6(30)$~efm$^2$ coefficient, or, to compare with Fig. 5 of Ref.~\cite{boso19}, $c/2a =c/M_0= 0.11(4)$. This alternative way also shows a significant deviation from isospin symmetry.

Deviations from the linearity of $M_\text{p}$ may be explained by isospin mixing of $T=0$ and $T=1$ states in the odd-odd system. However, isospin mixing alone cannot explain the observed change in collectivity in \nuc{70}{Kr}. The dramatic change in the magnitude of the \bev\ between \nuc{70}{Se} and \nuc{70}{Kr} suggests a change in deformation with larger deformation for \nuc{70}{Kr} than for its mirror nucleus.

The theoretical predictions for the $A=70$ triplet are summarized in Table~\ref{tab:theory} and also shown in the bottom panel of Fig.~\ref{fig:mats}.
\begin{table}[ht]
  \caption{Selected theoretical predictions for the $B(E2; 0^+_1\rightarrow 2^+_1)$ values of the $A=70$ triplet. For the shell model calculations effective charges $e_\text{n} = 0.5e$ and $e_\text{p} = 1.5e$ were used.}
  \begin{threeparttable}
    \begin{tabular*}{\columnwidth}{l@{\extracolsep{\fill}}lllrr}
      \hline
      \hline
      Method                & \multicolumn{3}{c}{\beup\ (e$^2$fm$^4$) } & Reference \\
                            & \nuc{70}{Kr} & \nuc{70}{Br} & \nuc{70}{Se} & \\
      \hline
      \hline
      Experiment            & 2726(451)\tnote{a} & 1455(159)\tnote{b}      & 1699(84) \tnote{c}    & \\
      \hline                                                             
      HFB-5DCH              & 3289         &              & 2767         & this work\\
      HFB-SCCM              & 4725         &              & 3450         & this work\\
      FRDM12                & 4725         & 4465         & 3715         & \cite{moeller16}\\
      GXPF1A~\cite{honma05} & 1910         & 1990         & 2075         & \\
      JUN45~\cite{honma09}  & 2325         & 2085         & 1885         & \\
      VAMPIR                & 2945         & 2630         & 2365         & \cite{petrovici18b}\\
      \hline
      \hline
    \end{tabular*}
    \label{tab:theory}
    \begin{tablenotes}\footnotesize
    \item[a] present work
    \item[b] weighted average of present work and Ref.~\cite{nichols14}
    \item[c] weighted average of present work and Refs.~\cite{ljungvall08,nichols14}
    \end{tablenotes}
  \end{threeparttable}
\end{table}
Few calculations have been performed for all three $A=70$ isotopes within the same theoretical framework. The Hartree-Fock-Bogoliubov-based (HFB) models \cite{delaroche10,rodriguez14} were only applied to the even-even nuclei. They predict shape coexistence between an oblate and a triaxial shape with very little difference in both \nuc{70}{Se} and \nuc{70}{Kr}, but generally too large \bevs.
The increase in deformation towards \nuc{70}{Kr} is more pronounced in the symmetry conserving configuration-mixing (SCCM) method~\cite{rodriguez14} than in the five-dimensional collective quadrupole Hamiltonian (5DCH) treatment to account for the configuration mixing.
The Finite-Range Droplet-Model (FRDM12)~\cite{moeller16} predicts only the ground-state deformation parameter $\beta_2$ on the mean-field level. The \bevs\ shown in Table~\ref{tab:theory} have been calculated assuming a simple rotor model. These calculations again over-estimate the absolute deformation, and show an increase of deformation towards \nuc{70}{Kr} similar to the SCCM.
Shell-model calculations with the GXPF1A~\cite{honma04,honma05} and JUN45~\cite{honma09} effective interactions were previously performed for \nuc{70}{Se} and \nuc{70}{Br} and reproduce the observed \bevs\ quite well \cite{nichols14}. We have extended these calculations to include \nuc{70}{Kr} and find a decreasing linear trend as expected from Eq.~\ref{eq:linmat} for the GXPF1A effective interaction, in contrast to our experimental findings. The results obtained with the JUN45 effective interaction show a positive trend. Both shell model calculations are able to reproduce the absolute magnitude of the \bevs\ rather well, but also fail to reproduce the strong increase observed in \nuc{70}{Kr} compared to \nuc{70}{Br}. It should be noted that the inclusion of isospin non-conserving terms into the interaction, that are commonly added to explain mirror energy differences~\cite{bentley07,bentley15}, have a negligible effect on the calculated \bevs.
Finally, several calculations have been published using the complex excited VAMPIR model~\cite{petrovici15,petrovici17,petrovici18,petrovici18b}. The published values vary considerably over time, demonstrating the difficulty to correctly describe these shape-changing nuclei.
The latest results show shape coexistence between oblate and prolate shapes with a moderate, continuous increase of the \bevs\ toward \nuc{70}{Kr}~\cite{petrovici18b}, again in contradiction to the experimental result. 
However, it is interesting to note that this model is the only one to predict a shape change along the isobars since the wave functions of low-lying yrast states in \nuc{70}{Kr} and \nuc{70}{Br} are dominated by prolate components, while the oblate component becomes more important in \nuc{70}{Se}. The latter is also consistent with the conclusions of \cite{ljungvall08}. 

In conclusion, while several calculations show a slight increase of the matrix element (and hence the deformation) from \nuc{70}{Se} to \nuc{70}{Kr} 
no calculation is able to describe the absolute $B(E2)$ values and the strong increase between the mirror nuclei \nuc{70}{Se} and \nuc{70}{Kr} observed experimentally. 

In summary, we have determined the \bevupn{1} for the $T_z=-1$ nucleus \nuc{70}{Kr} for the first time.
In addition, previously known \bevupn{1}\ values for the isobars \nuc{70}{Br} and \nuc{70}{Se} were confirmed.
The $A=70$ triplet is the heaviest one where the \bevs\ for all three members are experimentally determined. The \bev\ of \nuc{70}{Kr} is significantly larger than in the other members of the $T=1$ triplet \nuc{70}{Br} and \nuc{70}{Se}. Proton matrix elements for the triplet have been extracted from the \bevs, and they should exhibit a simple linear relation as a function of isospin $T_z$. The large value determined for \nuc{70}{Kr} deviates by $3\sigma$ from the extrapolation based on the other two nuclei. This suggests that a substantial shape change occurs between the oblate \nuc{70}{Se}~\cite{ljungvall08} and (presumably prolate) \nuc{70}{Kr}~\cite{petrovici18b}. None of the current nuclear structure models is able to explain the increase of the \bev\ determined in this work.

\acknowledgments
We would like to thank the RIKEN accelerator and BigRIPS teams for providing the high intensity beams.
This work has been supported by UK STFC under grant numbers ST/L005727/1 and ST/P003885/1, the Spanish Ministerio de Econom\'ia y Competitividad under grants FPA2011-24553 and FPA2014-52823-C2-1-P, the Program Severo Ochoa (SEV-2014-0398), and the Spanish MICINN under PGC2018-094583-B-I00.
K. W. acknowledges the support from the Spanish Ministerio de Econom\'ia y Competitividad RYC-2017-22007.
A. O. acknowledges the support from the European Research Council through the ERC Grant No. MINOS-258567.
\bibliography{draft}

\begin{thebibliography}{42}%
\makeatletter
\providecommand \@ifxundefined [1]{%
 \@ifx{#1\undefined}
}%
\providecommand \@ifnum [1]{%
 \ifnum #1\expandafter \@firstoftwo
 \else \expandafter \@secondoftwo
 \fi
}%
\providecommand \@ifx [1]{%
 \ifx #1\expandafter \@firstoftwo
 \else \expandafter \@secondoftwo
 \fi
}%
\providecommand \natexlab [1]{#1}%
\providecommand \enquote  [1]{``#1''}%
\providecommand \bibnamefont  [1]{#1}%
\providecommand \bibfnamefont [1]{#1}%
\providecommand \citenamefont [1]{#1}%
\providecommand \href@noop [0]{\@secondoftwo}%
\providecommand \href [0]{\begingroup \@sanitize@url \@href}%
\providecommand \@href[1]{\@@startlink{#1}\@@href}%
\providecommand \@@href[1]{\endgroup#1\@@endlink}%
\providecommand \@sanitize@url [0]{\catcode `\\12\catcode `\$12\catcode
  `\&12\catcode `\#12\catcode `\^12\catcode `\_12\catcode `\%12\relax}%
\providecommand \@@startlink[1]{}%
\providecommand \@@endlink[0]{}%
\providecommand \url  [0]{\begingroup\@sanitize@url \@url }%
\providecommand \@url [1]{\endgroup\@href {#1}{\urlprefix }}%
\providecommand \urlprefix  [0]{URL }%
\providecommand \Eprint [0]{\href }%
\providecommand \doibase [0]{http://dx.doi.org/}%
\providecommand \selectlanguage [0]{\@gobble}%
\providecommand \bibinfo  [0]{\@secondoftwo}%
\providecommand \bibfield  [0]{\@secondoftwo}%
\providecommand \translation [1]{[#1]}%
\providecommand \BibitemOpen [0]{}%
\providecommand \bibitemStop [0]{}%
\providecommand \bibitemNoStop [0]{.\EOS\space}%
\providecommand \EOS [0]{\spacefactor3000\relax}%
\providecommand \BibitemShut  [1]{\csname bibitem#1\endcsname}%
\let\auto@bib@innerbib\@empty
\bibitem [{\citenamefont {Heisenberg}(1932)}]{heisenberg32}%
  \BibitemOpen
  \bibfield  {author} {\bibinfo {author} {\bibfnamefont {W.}~\bibnamefont
  {Heisenberg}},\ }\href@noop {} {\bibfield  {journal} {\bibinfo  {journal} {Z.
  Phys.}\ }\textbf {\bibinfo {volume} {77}},\ \bibinfo {pages} {1} (\bibinfo
  {year} {1932})}\BibitemShut {NoStop}%
\bibitem [{\citenamefont {Miller}\ \emph {et~al.}(2006)\citenamefont {Miller},
  \citenamefont {Opper},\ and\ \citenamefont {Stephenson}}]{miller06}%
  \BibitemOpen
  \bibfield  {author} {\bibinfo {author} {\bibfnamefont {G.~A.}\ \bibnamefont
  {Miller}}, \bibinfo {author} {\bibfnamefont {A.~K.}\ \bibnamefont {Opper}}, \
  and\ \bibinfo {author} {\bibfnamefont {E.~J.}\ \bibnamefont {Stephenson}},\
  }\href {\doibase 10.1146/annurev.nucl.56.080805.140446} {\bibfield  {journal}
  {\bibinfo  {journal} {Ann. Rev. Nucl. Part. Sci.}\ }\textbf {\bibinfo
  {volume} {56}},\ \bibinfo {pages} {253} (\bibinfo {year} {2006})}\BibitemShut
  {NoStop}%
\bibitem [{\citenamefont {Wigner}(1937)}]{wigner37}%
  \BibitemOpen
  \bibfield  {author} {\bibinfo {author} {\bibfnamefont {E.}~\bibnamefont
  {Wigner}},\ }\href {\doibase 10.1103/PhysRev.51.106} {\bibfield  {journal}
  {\bibinfo  {journal} {Phys. Rev.}\ }\textbf {\bibinfo {volume} {51}},\
  \bibinfo {pages} {106} (\bibinfo {year} {1937})}\BibitemShut {NoStop}%
\bibitem [{\citenamefont {Hoff}\ \emph {et~al.}(2020)\citenamefont {Hoff},
  \citenamefont {Rogers}, \citenamefont {Wang}, \citenamefont {Bender},
  \citenamefont {Brandenburg}, \citenamefont {Childers}, \citenamefont {Clark},
  \citenamefont {Dombos}, \citenamefont {Doucet}, \citenamefont {Jin},
  \citenamefont {Lewis}, \citenamefont {Liddick}, \citenamefont {Lister},
  \citenamefont {Meisel}, \citenamefont {Morse}, \citenamefont {Nazarewicz},
  \citenamefont {Schatz}, \citenamefont {Schmidt}, \citenamefont {Soltesz},
  \citenamefont {Subedi},\ and\ \citenamefont {Waniganeththi}}]{hoff20}%
  \BibitemOpen
  \bibfield  {author} {\bibinfo {author} {\bibfnamefont {D.~E.~M.}\
  \bibnamefont {Hoff}}, \bibinfo {author} {\bibfnamefont {A.~M.}\ \bibnamefont
  {Rogers}}, \bibinfo {author} {\bibfnamefont {S.~M.}\ \bibnamefont {Wang}},
  \bibinfo {author} {\bibfnamefont {P.~C.}\ \bibnamefont {Bender}}, \bibinfo
  {author} {\bibfnamefont {K.}~\bibnamefont {Brandenburg}}, \bibinfo {author}
  {\bibfnamefont {K.}~\bibnamefont {Childers}}, \bibinfo {author}
  {\bibfnamefont {J.~A.}\ \bibnamefont {Clark}}, \bibinfo {author}
  {\bibfnamefont {A.~C.}\ \bibnamefont {Dombos}}, \bibinfo {author}
  {\bibfnamefont {E.~R.}\ \bibnamefont {Doucet}}, \bibinfo {author}
  {\bibfnamefont {S.}~\bibnamefont {Jin}}, \bibinfo {author} {\bibfnamefont
  {R.}~\bibnamefont {Lewis}}, \bibinfo {author} {\bibfnamefont {S.~N.}\
  \bibnamefont {Liddick}}, \bibinfo {author} {\bibfnamefont {C.~J.}\
  \bibnamefont {Lister}}, \bibinfo {author} {\bibfnamefont {Z.}~\bibnamefont
  {Meisel}}, \bibinfo {author} {\bibfnamefont {C.}~\bibnamefont {Morse}},
  \bibinfo {author} {\bibfnamefont {W.}~\bibnamefont {Nazarewicz}}, \bibinfo
  {author} {\bibfnamefont {H.}~\bibnamefont {Schatz}}, \bibinfo {author}
  {\bibfnamefont {K.}~\bibnamefont {Schmidt}}, \bibinfo {author} {\bibfnamefont
  {D.}~\bibnamefont {Soltesz}}, \bibinfo {author} {\bibfnamefont {S.~K.}\
  \bibnamefont {Subedi}}, \ and\ \bibinfo {author} {\bibfnamefont
  {S.}~\bibnamefont {Waniganeththi}},\ }\href
  {https://doi.org/10.1038/s41586-020-2123-1} {\bibfield  {journal} {\bibinfo
  {journal} {Nature}\ }\textbf {\bibinfo {volume} {580}},\ \bibinfo {pages}
  {52} (\bibinfo {year} {2020})}\BibitemShut {NoStop}%
\bibitem [{\citenamefont {Bentley}\ and\ \citenamefont
  {Lenzi}(2007)}]{bentley07}%
  \BibitemOpen
  \bibfield  {author} {\bibinfo {author} {\bibfnamefont {M.~A.}\ \bibnamefont
  {Bentley}}\ and\ \bibinfo {author} {\bibfnamefont {S.~M.}\ \bibnamefont
  {Lenzi}},\ }\href
  {http://www.sciencedirect.com/science/article/pii/S0146641006000743}
  {\bibfield  {journal} {\bibinfo  {journal} {Prog. Part. Nucl. Phys.}\
  }\textbf {\bibinfo {volume} {59}},\ \bibinfo {pages} {497} (\bibinfo {year}
  {2007})}\BibitemShut {NoStop}%
\bibitem [{\citenamefont {Zuker}\ \emph {et~al.}(2002)\citenamefont {Zuker},
  \citenamefont {Lenzi}, \citenamefont {Mart\'{\i}nez-Pinedo},\ and\
  \citenamefont {Poves}}]{zuker02}%
  \BibitemOpen
  \bibfield  {author} {\bibinfo {author} {\bibfnamefont {A.~P.}\ \bibnamefont
  {Zuker}}, \bibinfo {author} {\bibfnamefont {S.~M.}\ \bibnamefont {Lenzi}},
  \bibinfo {author} {\bibfnamefont {G.}~\bibnamefont {Mart\'{\i}nez-Pinedo}}, \
  and\ \bibinfo {author} {\bibfnamefont {A.}~\bibnamefont {Poves}},\ }\href
  {\doibase 10.1103/PhysRevLett.89.142502} {\bibfield  {journal} {\bibinfo
  {journal} {Phys. Rev. Lett.}\ }\textbf {\bibinfo {volume} {89}},\ \bibinfo
  {pages} {142502} (\bibinfo {year} {2002})}\BibitemShut {NoStop}%
\bibitem [{\citenamefont {Bernstein}\ \emph {et~al.}(1979)\citenamefont
  {Bernstein}, \citenamefont {Brown},\ and\ \citenamefont
  {Madsen}}]{bernstein79}%
  \BibitemOpen
  \bibfield  {author} {\bibinfo {author} {\bibfnamefont {A.~M.}\ \bibnamefont
  {Bernstein}}, \bibinfo {author} {\bibfnamefont {V.~R.}\ \bibnamefont
  {Brown}}, \ and\ \bibinfo {author} {\bibfnamefont {V.~A.}\ \bibnamefont
  {Madsen}},\ }\href {\doibase 10.1103/PhysRevLett.42.425} {\bibfield
  {journal} {\bibinfo  {journal} {Phys. Rev. Lett.}\ }\textbf {\bibinfo
  {volume} {42}},\ \bibinfo {pages} {425} (\bibinfo {year} {1979})}\BibitemShut
  {NoStop}%
\bibitem [{\citenamefont {Prados~Est\'evez}\ \emph {et~al.}(2007)\citenamefont
  {Prados~Est\'evez}, \citenamefont {Bruce}, \citenamefont {Taylor},
  \citenamefont {Amro}, \citenamefont {Beausang}, \citenamefont {Casten},
  \citenamefont {Ressler}, \citenamefont {Barton}, \citenamefont {Chandler},\
  and\ \citenamefont {Hammond}}]{pradosestevez07}%
  \BibitemOpen
  \bibfield  {author} {\bibinfo {author} {\bibfnamefont {F.~M.}\ \bibnamefont
  {Prados~Est\'evez}}, \bibinfo {author} {\bibfnamefont {A.~M.}\ \bibnamefont
  {Bruce}}, \bibinfo {author} {\bibfnamefont {M.~J.}\ \bibnamefont {Taylor}},
  \bibinfo {author} {\bibfnamefont {H.}~\bibnamefont {Amro}}, \bibinfo {author}
  {\bibfnamefont {C.~W.}\ \bibnamefont {Beausang}}, \bibinfo {author}
  {\bibfnamefont {R.~F.}\ \bibnamefont {Casten}}, \bibinfo {author}
  {\bibfnamefont {J.~J.}\ \bibnamefont {Ressler}}, \bibinfo {author}
  {\bibfnamefont {C.~J.}\ \bibnamefont {Barton}}, \bibinfo {author}
  {\bibfnamefont {C.}~\bibnamefont {Chandler}}, \ and\ \bibinfo {author}
  {\bibfnamefont {G.}~\bibnamefont {Hammond}},\ }\href {\doibase
  10.1103/PhysRevC.75.014309} {\bibfield  {journal} {\bibinfo  {journal} {Phys.
  Rev. C}\ }\textbf {\bibinfo {volume} {75}},\ \bibinfo {pages} {014309}
  (\bibinfo {year} {2007})}\BibitemShut {NoStop}%
\bibitem [{\citenamefont {Boso}\ \emph {et~al.}(2019)\citenamefont {Boso},
  \citenamefont {Milne}, \citenamefont {Bentley}, \citenamefont {Recchia},
  \citenamefont {Lenzi}, \citenamefont {Rudolph}, \citenamefont {Labiche},
  \citenamefont {Pereira-Lopez}, \citenamefont {Afara}, \citenamefont {Ameil},
  \citenamefont {Arici}, \citenamefont {Aydin}, \citenamefont {Axiotis},
  \citenamefont {Barrientos}, \citenamefont {Benzoni}, \citenamefont
  {Birkenbach}, \citenamefont {Boston}, \citenamefont {Boston}, \citenamefont
  {Boutachkov}, \citenamefont {Bracco}, \citenamefont {Bruce}, \citenamefont
  {Bruyneel}, \citenamefont {Cederwall}, \citenamefont {Clement}, \citenamefont
  {Cortes}, \citenamefont {Cullen}, \citenamefont {D\'esesquelles},
  \citenamefont {Dombr\'adi}, \citenamefont {Domingo-Pardo}, \citenamefont
  {Eberth}, \citenamefont {Fahlander}, \citenamefont {Gelain}, \citenamefont
  {Gonz\'alez}, \citenamefont {John}, \citenamefont {Gerl}, \citenamefont
  {Golubev}, \citenamefont {G\'orska}, \citenamefont {Gottardo}, \citenamefont
  {Grahn}, \citenamefont {Grassi}, \citenamefont {Habermann}, \citenamefont
  {Harkness-Brennan}, \citenamefont {Henry}, \citenamefont {Hess},
  \citenamefont {Kojouharov}, \citenamefont {Korten}, \citenamefont
  {Lalovi\'c}, \citenamefont {Lettmann}, \citenamefont {Lizarazo},
  \citenamefont {Louchart-Henning}, \citenamefont {Menegazzo}, \citenamefont
  {Mengoni}, \citenamefont {Merchan}, \citenamefont {Michelagnoli},
  \citenamefont {Million}, \citenamefont {Modamio}, \citenamefont {Moeller},
  \citenamefont {Napoli}, \citenamefont {Nyberg}, \citenamefont {Singh]},
  \citenamefont {Pai}, \citenamefont {Pietralla}, \citenamefont {Pietri},
  \citenamefont {Podolyak}, \citenamefont {Vidal]}, \citenamefont {Pullia},
  \citenamefont {Ralet}, \citenamefont {Rainovski}, \citenamefont {Reese},
  \citenamefont {Reiter}, \citenamefont {Salsac}, \citenamefont {Sanchis},
  \citenamefont {Sarmiento}, \citenamefont {Schaffner}, \citenamefont
  {Scruton}, \citenamefont {Singh}, \citenamefont {Stahl}, \citenamefont
  {Uthayakumaar}, \citenamefont {Valiente-Dob\'on},\ and\ \citenamefont
  {Wieland}}]{boso19}%
  \BibitemOpen
  \bibfield  {author} {\bibinfo {author} {\bibfnamefont {A.}~\bibnamefont
  {Boso}}, \bibinfo {author} {\bibfnamefont {S.}~\bibnamefont {Milne}},
  \bibinfo {author} {\bibfnamefont {M.}~\bibnamefont {Bentley}}, \bibinfo
  {author} {\bibfnamefont {F.}~\bibnamefont {Recchia}}, \bibinfo {author}
  {\bibfnamefont {S.}~\bibnamefont {Lenzi}}, \bibinfo {author} {\bibfnamefont
  {D.}~\bibnamefont {Rudolph}}, \bibinfo {author} {\bibfnamefont
  {M.}~\bibnamefont {Labiche}}, \bibinfo {author} {\bibfnamefont
  {X.}~\bibnamefont {Pereira-Lopez}}, \bibinfo {author} {\bibfnamefont
  {S.}~\bibnamefont {Afara}}, \bibinfo {author} {\bibfnamefont
  {F.}~\bibnamefont {Ameil}}, \bibinfo {author} {\bibfnamefont
  {T.}~\bibnamefont {Arici}}, \bibinfo {author} {\bibfnamefont
  {S.}~\bibnamefont {Aydin}}, \bibinfo {author} {\bibfnamefont
  {M.}~\bibnamefont {Axiotis}}, \bibinfo {author} {\bibfnamefont
  {D.}~\bibnamefont {Barrientos}}, \bibinfo {author} {\bibfnamefont
  {G.}~\bibnamefont {Benzoni}}, \bibinfo {author} {\bibfnamefont
  {B.}~\bibnamefont {Birkenbach}}, \bibinfo {author} {\bibfnamefont
  {A.}~\bibnamefont {Boston}}, \bibinfo {author} {\bibfnamefont
  {H.}~\bibnamefont {Boston}}, \bibinfo {author} {\bibfnamefont
  {P.}~\bibnamefont {Boutachkov}}, \bibinfo {author} {\bibfnamefont
  {A.}~\bibnamefont {Bracco}}, \bibinfo {author} {\bibfnamefont
  {A.}~\bibnamefont {Bruce}}, \bibinfo {author} {\bibfnamefont
  {B.}~\bibnamefont {Bruyneel}}, \bibinfo {author} {\bibfnamefont
  {B.}~\bibnamefont {Cederwall}}, \bibinfo {author} {\bibfnamefont
  {E.}~\bibnamefont {Clement}}, \bibinfo {author} {\bibfnamefont
  {M.}~\bibnamefont {Cortes}}, \bibinfo {author} {\bibfnamefont
  {D.}~\bibnamefont {Cullen}}, \bibinfo {author} {\bibfnamefont
  {P.}~\bibnamefont {D\'esesquelles}}, \bibinfo {author} {\bibfnamefont
  {Z.}~\bibnamefont {Dombr\'adi}}, \bibinfo {author} {\bibfnamefont
  {C.}~\bibnamefont {Domingo-Pardo}}, \bibinfo {author} {\bibfnamefont
  {J.}~\bibnamefont {Eberth}}, \bibinfo {author} {\bibfnamefont
  {C.}~\bibnamefont {Fahlander}}, \bibinfo {author} {\bibfnamefont
  {M.}~\bibnamefont {Gelain}}, \bibinfo {author} {\bibfnamefont
  {V.}~\bibnamefont {Gonz\'alez}}, \bibinfo {author} {\bibfnamefont
  {P.}~\bibnamefont {John}}, \bibinfo {author} {\bibfnamefont {J.}~\bibnamefont
  {Gerl}}, \bibinfo {author} {\bibfnamefont {P.}~\bibnamefont {Golubev}},
  \bibinfo {author} {\bibfnamefont {M.}~\bibnamefont {G\'orska}}, \bibinfo
  {author} {\bibfnamefont {A.}~\bibnamefont {Gottardo}}, \bibinfo {author}
  {\bibfnamefont {T.}~\bibnamefont {Grahn}}, \bibinfo {author} {\bibfnamefont
  {L.}~\bibnamefont {Grassi}}, \bibinfo {author} {\bibfnamefont
  {T.}~\bibnamefont {Habermann}}, \bibinfo {author} {\bibfnamefont
  {L.}~\bibnamefont {Harkness-Brennan}}, \bibinfo {author} {\bibfnamefont
  {T.}~\bibnamefont {Henry}}, \bibinfo {author} {\bibfnamefont
  {H.}~\bibnamefont {Hess}}, \bibinfo {author} {\bibfnamefont {I.}~\bibnamefont
  {Kojouharov}}, \bibinfo {author} {\bibfnamefont {W.}~\bibnamefont {Korten}},
  \bibinfo {author} {\bibfnamefont {N.}~\bibnamefont {Lalovi\'c}}, \bibinfo
  {author} {\bibfnamefont {M.}~\bibnamefont {Lettmann}}, \bibinfo {author}
  {\bibfnamefont {C.}~\bibnamefont {Lizarazo}}, \bibinfo {author}
  {\bibfnamefont {C.}~\bibnamefont {Louchart-Henning}}, \bibinfo {author}
  {\bibfnamefont {R.}~\bibnamefont {Menegazzo}}, \bibinfo {author}
  {\bibfnamefont {D.}~\bibnamefont {Mengoni}}, \bibinfo {author} {\bibfnamefont
  {E.}~\bibnamefont {Merchan}}, \bibinfo {author} {\bibfnamefont
  {C.}~\bibnamefont {Michelagnoli}}, \bibinfo {author} {\bibfnamefont
  {B.}~\bibnamefont {Million}}, \bibinfo {author} {\bibfnamefont
  {V.}~\bibnamefont {Modamio}}, \bibinfo {author} {\bibfnamefont
  {T.}~\bibnamefont {Moeller}}, \bibinfo {author} {\bibfnamefont
  {D.}~\bibnamefont {Napoli}}, \bibinfo {author} {\bibfnamefont
  {J.}~\bibnamefont {Nyberg}}, \bibinfo {author} {\bibfnamefont {B.~N.}\
  \bibnamefont {Singh]}}, \bibinfo {author} {\bibfnamefont {H.}~\bibnamefont
  {Pai}}, \bibinfo {author} {\bibfnamefont {N.}~\bibnamefont {Pietralla}},
  \bibinfo {author} {\bibfnamefont {S.}~\bibnamefont {Pietri}}, \bibinfo
  {author} {\bibfnamefont {Z.}~\bibnamefont {Podolyak}}, \bibinfo {author}
  {\bibfnamefont {R.~P.}\ \bibnamefont {Vidal]}}, \bibinfo {author}
  {\bibfnamefont {A.}~\bibnamefont {Pullia}}, \bibinfo {author} {\bibfnamefont
  {D.}~\bibnamefont {Ralet}}, \bibinfo {author} {\bibfnamefont
  {G.}~\bibnamefont {Rainovski}}, \bibinfo {author} {\bibfnamefont
  {M.}~\bibnamefont {Reese}}, \bibinfo {author} {\bibfnamefont
  {P.}~\bibnamefont {Reiter}}, \bibinfo {author} {\bibfnamefont
  {M.}~\bibnamefont {Salsac}}, \bibinfo {author} {\bibfnamefont
  {E.}~\bibnamefont {Sanchis}}, \bibinfo {author} {\bibfnamefont
  {L.}~\bibnamefont {Sarmiento}}, \bibinfo {author} {\bibfnamefont
  {H.}~\bibnamefont {Schaffner}}, \bibinfo {author} {\bibfnamefont
  {L.}~\bibnamefont {Scruton}}, \bibinfo {author} {\bibfnamefont
  {P.}~\bibnamefont {Singh}}, \bibinfo {author} {\bibfnamefont
  {C.}~\bibnamefont {Stahl}}, \bibinfo {author} {\bibfnamefont
  {S.}~\bibnamefont {Uthayakumaar}}, \bibinfo {author} {\bibfnamefont
  {J.}~\bibnamefont {Valiente-Dob\'on}}, \ and\ \bibinfo {author}
  {\bibfnamefont {O.}~\bibnamefont {Wieland}},\ }\href {\doibase
  https://doi.org/10.1016/j.physletb.2019.134835} {\bibfield  {journal}
  {\bibinfo  {journal} {Phys. Lett. B}\ }\textbf {\bibinfo {volume} {797}},\
  \bibinfo {pages} {134835} (\bibinfo {year} {2019})}\BibitemShut {NoStop}%
\bibitem [{\citenamefont {Giles}\ \emph {et~al.}(2019)\citenamefont {Giles},
  \citenamefont {Nara~Singh}, \citenamefont {Barber}, \citenamefont {Cullen},
  \citenamefont {Mallaburn}, \citenamefont {Beckers}, \citenamefont {Blazhev},
  \citenamefont {Braunroth}, \citenamefont {Dewald}, \citenamefont {Fransen},
  \citenamefont {Goldkuhle}, \citenamefont {Jolie}, \citenamefont {Mammes},
  \citenamefont {M\"uller-Gatermann}, \citenamefont {W\"olk}, \citenamefont
  {Zell}, \citenamefont {Lenzi},\ and\ \citenamefont {Poves}}]{giles19}%
  \BibitemOpen
  \bibfield  {author} {\bibinfo {author} {\bibfnamefont {M.~M.}\ \bibnamefont
  {Giles}}, \bibinfo {author} {\bibfnamefont {B.~S.}\ \bibnamefont
  {Nara~Singh}}, \bibinfo {author} {\bibfnamefont {L.}~\bibnamefont {Barber}},
  \bibinfo {author} {\bibfnamefont {D.~M.}\ \bibnamefont {Cullen}}, \bibinfo
  {author} {\bibfnamefont {M.~J.}\ \bibnamefont {Mallaburn}}, \bibinfo {author}
  {\bibfnamefont {M.}~\bibnamefont {Beckers}}, \bibinfo {author} {\bibfnamefont
  {A.}~\bibnamefont {Blazhev}}, \bibinfo {author} {\bibfnamefont
  {T.}~\bibnamefont {Braunroth}}, \bibinfo {author} {\bibfnamefont
  {A.}~\bibnamefont {Dewald}}, \bibinfo {author} {\bibfnamefont
  {C.}~\bibnamefont {Fransen}}, \bibinfo {author} {\bibfnamefont
  {A.}~\bibnamefont {Goldkuhle}}, \bibinfo {author} {\bibfnamefont
  {J.}~\bibnamefont {Jolie}}, \bibinfo {author} {\bibfnamefont
  {F.}~\bibnamefont {Mammes}}, \bibinfo {author} {\bibfnamefont
  {C.}~\bibnamefont {M\"uller-Gatermann}}, \bibinfo {author} {\bibfnamefont
  {D.}~\bibnamefont {W\"olk}}, \bibinfo {author} {\bibfnamefont {K.~O.}\
  \bibnamefont {Zell}}, \bibinfo {author} {\bibfnamefont {S.~M.}\ \bibnamefont
  {Lenzi}}, \ and\ \bibinfo {author} {\bibfnamefont {A.}~\bibnamefont
  {Poves}},\ }\href {\doibase 10.1103/PhysRevC.99.044317} {\bibfield  {journal}
  {\bibinfo  {journal} {Phys. Rev. C}\ }\textbf {\bibinfo {volume} {99}},\
  \bibinfo {pages} {044317} (\bibinfo {year} {2019})}\BibitemShut {NoStop}%
\bibitem [{\citenamefont {de~Angelis}\ \emph {et~al.}(2001)\citenamefont
  {de~Angelis}, \citenamefont {Martinez}, \citenamefont {Gadea}, \citenamefont
  {Marginean}, \citenamefont {Farnea}, \citenamefont {Maglione}, \citenamefont
  {Lenzi}, \citenamefont {Gelletly}, \citenamefont {Ur}, \citenamefont
  {Napoli}, \citenamefont {Kroell}, \citenamefont {Lunardi}, \citenamefont
  {Rubio}, \citenamefont {Axiotis}, \citenamefont {Bazzacco}, \citenamefont
  {Sona}, \citenamefont {Bizzeti}, \citenamefont {Bednarczyk}, \citenamefont
  {Bracco}, \citenamefont {Brandolini}, \citenamefont {Camera}, \citenamefont
  {Curien}, \citenamefont {Poli}, \citenamefont {Dorvaux}, \citenamefont
  {Eberth}, \citenamefont {Grawe}, \citenamefont {Menegazzo}, \citenamefont
  {Nardelli}, \citenamefont {Nyberg}, \citenamefont {Pavan}, \citenamefont
  {Quintana}, \citenamefont {Alvarez}, \citenamefont {Spolaore}, \citenamefont
  {Steinhart}, \citenamefont {Stefanescu}, \citenamefont {Thelen},\ and\
  \citenamefont {Venturelli}}]{deangelis01}%
  \BibitemOpen
  \bibfield  {author} {\bibinfo {author} {\bibfnamefont {G.}~\bibnamefont
  {de~Angelis}}, \bibinfo {author} {\bibfnamefont {T.}~\bibnamefont
  {Martinez}}, \bibinfo {author} {\bibfnamefont {A.}~\bibnamefont {Gadea}},
  \bibinfo {author} {\bibfnamefont {N.}~\bibnamefont {Marginean}}, \bibinfo
  {author} {\bibfnamefont {E.}~\bibnamefont {Farnea}}, \bibinfo {author}
  {\bibfnamefont {E.}~\bibnamefont {Maglione}}, \bibinfo {author}
  {\bibfnamefont {S.}~\bibnamefont {Lenzi}}, \bibinfo {author} {\bibfnamefont
  {W.}~\bibnamefont {Gelletly}}, \bibinfo {author} {\bibfnamefont
  {C.}~\bibnamefont {Ur}}, \bibinfo {author} {\bibfnamefont {D.}~\bibnamefont
  {Napoli}}, \bibinfo {author} {\bibfnamefont {T.}~\bibnamefont {Kroell}},
  \bibinfo {author} {\bibfnamefont {S.}~\bibnamefont {Lunardi}}, \bibinfo
  {author} {\bibfnamefont {B.}~\bibnamefont {Rubio}}, \bibinfo {author}
  {\bibfnamefont {M.}~\bibnamefont {Axiotis}}, \bibinfo {author} {\bibfnamefont
  {D.}~\bibnamefont {Bazzacco}}, \bibinfo {author} {\bibfnamefont {A.~B.}\
  \bibnamefont {Sona}}, \bibinfo {author} {\bibfnamefont {P.}~\bibnamefont
  {Bizzeti}}, \bibinfo {author} {\bibfnamefont {P.}~\bibnamefont {Bednarczyk}},
  \bibinfo {author} {\bibfnamefont {A.}~\bibnamefont {Bracco}}, \bibinfo
  {author} {\bibfnamefont {F.}~\bibnamefont {Brandolini}}, \bibinfo {author}
  {\bibfnamefont {F.}~\bibnamefont {Camera}}, \bibinfo {author} {\bibfnamefont
  {D.}~\bibnamefont {Curien}}, \bibinfo {author} {\bibfnamefont {M.~D.}\
  \bibnamefont {Poli}}, \bibinfo {author} {\bibfnamefont {O.}~\bibnamefont
  {Dorvaux}}, \bibinfo {author} {\bibfnamefont {J.}~\bibnamefont {Eberth}},
  \bibinfo {author} {\bibfnamefont {H.}~\bibnamefont {Grawe}}, \bibinfo
  {author} {\bibfnamefont {R.}~\bibnamefont {Menegazzo}}, \bibinfo {author}
  {\bibfnamefont {G.}~\bibnamefont {Nardelli}}, \bibinfo {author}
  {\bibfnamefont {J.}~\bibnamefont {Nyberg}}, \bibinfo {author} {\bibfnamefont
  {P.}~\bibnamefont {Pavan}}, \bibinfo {author} {\bibfnamefont
  {B.}~\bibnamefont {Quintana}}, \bibinfo {author} {\bibfnamefont {C.~R.}\
  \bibnamefont {Alvarez}}, \bibinfo {author} {\bibfnamefont {P.}~\bibnamefont
  {Spolaore}}, \bibinfo {author} {\bibfnamefont {T.}~\bibnamefont {Steinhart}},
  \bibinfo {author} {\bibfnamefont {I.}~\bibnamefont {Stefanescu}}, \bibinfo
  {author} {\bibfnamefont {O.}~\bibnamefont {Thelen}}, \ and\ \bibinfo {author}
  {\bibfnamefont {R.}~\bibnamefont {Venturelli}},\ }\href {\doibase
  https://doi.org/10.1007/s100500170038} {\bibfield  {journal} {\bibinfo
  {journal} {Eur. Phys. Jour. A}\ }\textbf {\bibinfo {volume} {12}},\ \bibinfo
  {pages} {51} (\bibinfo {year} {2001})}\BibitemShut {NoStop}%
\bibitem [{\citenamefont {Nara~Singh}\ \emph {et~al.}(2007)\citenamefont
  {Nara~Singh}, \citenamefont {Steer}, \citenamefont {Jenkins}, \citenamefont
  {Wadsworth}, \citenamefont {Bentley}, \citenamefont {Davies}, \citenamefont
  {Glover}, \citenamefont {Pattabiraman}, \citenamefont {Lister}, \citenamefont
  {Grahn}, \citenamefont {Greenlees}, \citenamefont {Jones}, \citenamefont
  {Julin}, \citenamefont {Juutinen}, \citenamefont {Leino}, \citenamefont
  {Nyman}, \citenamefont {Pakarinen}, \citenamefont {Rahkila}, \citenamefont
  {Sar\'en}, \citenamefont {Scholey}, \citenamefont {Sorri}, \citenamefont
  {Uusitalo}, \citenamefont {Butler}, \citenamefont {Dimmock}, \citenamefont
  {Joss}, \citenamefont {Thomson}, \citenamefont {Cederwall}, \citenamefont
  {Hadinia},\ and\ \citenamefont {Sandzelius}}]{narasingh07}%
  \BibitemOpen
  \bibfield  {author} {\bibinfo {author} {\bibfnamefont {B.~S.}\ \bibnamefont
  {Nara~Singh}}, \bibinfo {author} {\bibfnamefont {A.~N.}\ \bibnamefont
  {Steer}}, \bibinfo {author} {\bibfnamefont {D.~G.}\ \bibnamefont {Jenkins}},
  \bibinfo {author} {\bibfnamefont {R.}~\bibnamefont {Wadsworth}}, \bibinfo
  {author} {\bibfnamefont {M.~A.}\ \bibnamefont {Bentley}}, \bibinfo {author}
  {\bibfnamefont {P.~J.}\ \bibnamefont {Davies}}, \bibinfo {author}
  {\bibfnamefont {R.}~\bibnamefont {Glover}}, \bibinfo {author} {\bibfnamefont
  {N.~S.}\ \bibnamefont {Pattabiraman}}, \bibinfo {author} {\bibfnamefont
  {C.~J.}\ \bibnamefont {Lister}}, \bibinfo {author} {\bibfnamefont
  {T.}~\bibnamefont {Grahn}}, \bibinfo {author} {\bibfnamefont {P.~T.}\
  \bibnamefont {Greenlees}}, \bibinfo {author} {\bibfnamefont {P.}~\bibnamefont
  {Jones}}, \bibinfo {author} {\bibfnamefont {R.}~\bibnamefont {Julin}},
  \bibinfo {author} {\bibfnamefont {S.}~\bibnamefont {Juutinen}}, \bibinfo
  {author} {\bibfnamefont {M.}~\bibnamefont {Leino}}, \bibinfo {author}
  {\bibfnamefont {M.}~\bibnamefont {Nyman}}, \bibinfo {author} {\bibfnamefont
  {J.}~\bibnamefont {Pakarinen}}, \bibinfo {author} {\bibfnamefont
  {P.}~\bibnamefont {Rahkila}}, \bibinfo {author} {\bibfnamefont
  {J.}~\bibnamefont {Sar\'en}}, \bibinfo {author} {\bibfnamefont
  {C.}~\bibnamefont {Scholey}}, \bibinfo {author} {\bibfnamefont
  {J.}~\bibnamefont {Sorri}}, \bibinfo {author} {\bibfnamefont
  {J.}~\bibnamefont {Uusitalo}}, \bibinfo {author} {\bibfnamefont {P.~A.}\
  \bibnamefont {Butler}}, \bibinfo {author} {\bibfnamefont {M.}~\bibnamefont
  {Dimmock}}, \bibinfo {author} {\bibfnamefont {D.~T.}\ \bibnamefont {Joss}},
  \bibinfo {author} {\bibfnamefont {J.}~\bibnamefont {Thomson}}, \bibinfo
  {author} {\bibfnamefont {B.}~\bibnamefont {Cederwall}}, \bibinfo {author}
  {\bibfnamefont {B.}~\bibnamefont {Hadinia}}, \ and\ \bibinfo {author}
  {\bibfnamefont {M.}~\bibnamefont {Sandzelius}},\ }\href {\doibase
  10.1103/PhysRevC.75.061301} {\bibfield  {journal} {\bibinfo  {journal} {Phys.
  Rev. C}\ }\textbf {\bibinfo {volume} {75}},\ \bibinfo {pages} {061301}
  (\bibinfo {year} {2007})}\BibitemShut {NoStop}%
\bibitem [{\citenamefont {Debenham}\ \emph {et~al.}(2016)\citenamefont
  {Debenham}, \citenamefont {Bentley}, \citenamefont {Davies}, \citenamefont
  {Haylett}, \citenamefont {Jenkins}, \citenamefont {Joshi}, \citenamefont
  {Sinclair}, \citenamefont {Wadsworth}, \citenamefont {Ruotsalainen},
  \citenamefont {Henderson}, \citenamefont {Kaneko}, \citenamefont {Auranen},
  \citenamefont {Badran}, \citenamefont {Grahn}, \citenamefont {Greenlees},
  \citenamefont {Herza\'a\ifmmode~\check{n}\else \v{n}\fi{}}, \citenamefont
  {Jakobsson}, \citenamefont {Konki}, \citenamefont {Julin}, \citenamefont
  {Juutinen}, \citenamefont {Leino}, \citenamefont {Sorri}, \citenamefont
  {Pakarinen}, \citenamefont {Papadakis}, \citenamefont {Peura}, \citenamefont
  {Partanen}, \citenamefont {Rahkila}, \citenamefont {Sandzelius},
  \citenamefont {Sar\'en}, \citenamefont {Scholey}, \citenamefont {Stolze},
  \citenamefont {Uusitalo}, \citenamefont {David}, \citenamefont {de~Angelis},
  \citenamefont {Korten}, \citenamefont {Lotay}, \citenamefont {Mallaburn},\
  and\ \citenamefont {Sahin}}]{debenham16}%
  \BibitemOpen
  \bibfield  {author} {\bibinfo {author} {\bibfnamefont {D.~M.}\ \bibnamefont
  {Debenham}}, \bibinfo {author} {\bibfnamefont {M.~A.}\ \bibnamefont
  {Bentley}}, \bibinfo {author} {\bibfnamefont {P.~J.}\ \bibnamefont {Davies}},
  \bibinfo {author} {\bibfnamefont {T.}~\bibnamefont {Haylett}}, \bibinfo
  {author} {\bibfnamefont {D.~G.}\ \bibnamefont {Jenkins}}, \bibinfo {author}
  {\bibfnamefont {P.}~\bibnamefont {Joshi}}, \bibinfo {author} {\bibfnamefont
  {L.~F.}\ \bibnamefont {Sinclair}}, \bibinfo {author} {\bibfnamefont
  {R.}~\bibnamefont {Wadsworth}}, \bibinfo {author} {\bibfnamefont
  {P.}~\bibnamefont {Ruotsalainen}}, \bibinfo {author} {\bibfnamefont
  {J.}~\bibnamefont {Henderson}}, \bibinfo {author} {\bibfnamefont
  {K.}~\bibnamefont {Kaneko}}, \bibinfo {author} {\bibfnamefont
  {K.}~\bibnamefont {Auranen}}, \bibinfo {author} {\bibfnamefont
  {H.}~\bibnamefont {Badran}}, \bibinfo {author} {\bibfnamefont
  {T.}~\bibnamefont {Grahn}}, \bibinfo {author} {\bibfnamefont
  {P.}~\bibnamefont {Greenlees}}, \bibinfo {author} {\bibfnamefont
  {A.}~\bibnamefont {Herza\'a\ifmmode~\check{n}\else \v{n}\fi{}}}, \bibinfo
  {author} {\bibfnamefont {U.}~\bibnamefont {Jakobsson}}, \bibinfo {author}
  {\bibfnamefont {J.}~\bibnamefont {Konki}}, \bibinfo {author} {\bibfnamefont
  {R.}~\bibnamefont {Julin}}, \bibinfo {author} {\bibfnamefont
  {S.}~\bibnamefont {Juutinen}}, \bibinfo {author} {\bibfnamefont
  {M.}~\bibnamefont {Leino}}, \bibinfo {author} {\bibfnamefont
  {J.}~\bibnamefont {Sorri}}, \bibinfo {author} {\bibfnamefont
  {J.}~\bibnamefont {Pakarinen}}, \bibinfo {author} {\bibfnamefont
  {P.}~\bibnamefont {Papadakis}}, \bibinfo {author} {\bibfnamefont
  {P.}~\bibnamefont {Peura}}, \bibinfo {author} {\bibfnamefont
  {J.}~\bibnamefont {Partanen}}, \bibinfo {author} {\bibfnamefont
  {P.}~\bibnamefont {Rahkila}}, \bibinfo {author} {\bibfnamefont
  {M.}~\bibnamefont {Sandzelius}}, \bibinfo {author} {\bibfnamefont
  {J.}~\bibnamefont {Sar\'en}}, \bibinfo {author} {\bibfnamefont
  {C.}~\bibnamefont {Scholey}}, \bibinfo {author} {\bibfnamefont
  {S.}~\bibnamefont {Stolze}}, \bibinfo {author} {\bibfnamefont
  {J.}~\bibnamefont {Uusitalo}}, \bibinfo {author} {\bibfnamefont {H.~M.}\
  \bibnamefont {David}}, \bibinfo {author} {\bibfnamefont {G.}~\bibnamefont
  {de~Angelis}}, \bibinfo {author} {\bibfnamefont {W.}~\bibnamefont {Korten}},
  \bibinfo {author} {\bibfnamefont {G.}~\bibnamefont {Lotay}}, \bibinfo
  {author} {\bibfnamefont {M.}~\bibnamefont {Mallaburn}}, \ and\ \bibinfo
  {author} {\bibfnamefont {E.}~\bibnamefont {Sahin}},\ }\href {\doibase
  10.1103/PhysRevC.94.054311} {\bibfield  {journal} {\bibinfo  {journal} {Phys.
  Rev. C}\ }\textbf {\bibinfo {volume} {94}},\ \bibinfo {pages} {054311}
  (\bibinfo {year} {2016})}\BibitemShut {NoStop}%
\bibitem [{\citenamefont {Wimmer}\ \emph {et~al.}(2018)\citenamefont {Wimmer},
  \citenamefont {Korten}, \citenamefont {Arici}, \citenamefont {Doornenbal},
  \citenamefont {Aguilera}, \citenamefont {Algora}, \citenamefont {Ando},
  \citenamefont {Baba}, \citenamefont {Blank}, \citenamefont {Boso},
  \citenamefont {Chen}, \citenamefont {Corsi}, \citenamefont {Davies},
  \citenamefont {de~Angelis}, \citenamefont {de~France}, \citenamefont
  {Doherty}, \citenamefont {Gerl}, \citenamefont {Gernh\"auser}, \citenamefont
  {Jenkins}, \citenamefont {Koyama}, \citenamefont {Motobayashi}, \citenamefont
  {Nagamine}, \citenamefont {Niikura}, \citenamefont {Obertelli}, \citenamefont
  {Lubos}, \citenamefont {Rubio}, \citenamefont {Sahin}, \citenamefont {Saito},
  \citenamefont {Sakurai}, \citenamefont {Sinclair}, \citenamefont
  {Steppenbeck}, \citenamefont {Taniuchi}, \citenamefont {Wadsworth},\ and\
  \citenamefont {Zielinska}}]{wimmer18}%
  \BibitemOpen
  \bibfield  {author} {\bibinfo {author} {\bibfnamefont {K.}~\bibnamefont
  {Wimmer}}, \bibinfo {author} {\bibfnamefont {W.}~\bibnamefont {Korten}},
  \bibinfo {author} {\bibfnamefont {T.}~\bibnamefont {Arici}}, \bibinfo
  {author} {\bibfnamefont {P.}~\bibnamefont {Doornenbal}}, \bibinfo {author}
  {\bibfnamefont {P.}~\bibnamefont {Aguilera}}, \bibinfo {author}
  {\bibfnamefont {A.}~\bibnamefont {Algora}}, \bibinfo {author} {\bibfnamefont
  {T.}~\bibnamefont {Ando}}, \bibinfo {author} {\bibfnamefont {H.}~\bibnamefont
  {Baba}}, \bibinfo {author} {\bibfnamefont {B.}~\bibnamefont {Blank}},
  \bibinfo {author} {\bibfnamefont {A.}~\bibnamefont {Boso}}, \bibinfo {author}
  {\bibfnamefont {S.}~\bibnamefont {Chen}}, \bibinfo {author} {\bibfnamefont
  {A.}~\bibnamefont {Corsi}}, \bibinfo {author} {\bibfnamefont
  {P.}~\bibnamefont {Davies}}, \bibinfo {author} {\bibfnamefont
  {G.}~\bibnamefont {de~Angelis}}, \bibinfo {author} {\bibfnamefont
  {G.}~\bibnamefont {de~France}}, \bibinfo {author} {\bibfnamefont
  {D.}~\bibnamefont {Doherty}}, \bibinfo {author} {\bibfnamefont
  {J.}~\bibnamefont {Gerl}}, \bibinfo {author} {\bibfnamefont {R.}~\bibnamefont
  {Gernh\"auser}}, \bibinfo {author} {\bibfnamefont {D.}~\bibnamefont
  {Jenkins}}, \bibinfo {author} {\bibfnamefont {S.}~\bibnamefont {Koyama}},
  \bibinfo {author} {\bibfnamefont {T.}~\bibnamefont {Motobayashi}}, \bibinfo
  {author} {\bibfnamefont {S.}~\bibnamefont {Nagamine}}, \bibinfo {author}
  {\bibfnamefont {M.}~\bibnamefont {Niikura}}, \bibinfo {author} {\bibfnamefont
  {A.}~\bibnamefont {Obertelli}}, \bibinfo {author} {\bibfnamefont
  {D.}~\bibnamefont {Lubos}}, \bibinfo {author} {\bibfnamefont
  {B.}~\bibnamefont {Rubio}}, \bibinfo {author} {\bibfnamefont
  {E.}~\bibnamefont {Sahin}}, \bibinfo {author} {\bibfnamefont
  {T.}~\bibnamefont {Saito}}, \bibinfo {author} {\bibfnamefont
  {H.}~\bibnamefont {Sakurai}}, \bibinfo {author} {\bibfnamefont
  {L.}~\bibnamefont {Sinclair}}, \bibinfo {author} {\bibfnamefont
  {D.}~\bibnamefont {Steppenbeck}}, \bibinfo {author} {\bibfnamefont
  {R.}~\bibnamefont {Taniuchi}}, \bibinfo {author} {\bibfnamefont
  {R.}~\bibnamefont {Wadsworth}}, \ and\ \bibinfo {author} {\bibfnamefont
  {M.}~\bibnamefont {Zielinska}},\ }\href {\doibase
  https://doi.org/10.1016/j.physletb.2018.07.067} {\bibfield  {journal}
  {\bibinfo  {journal} {Phys. Lett. B}\ }\textbf {\bibinfo {volume} {785}},\
  \bibinfo {pages} {441} (\bibinfo {year} {2018})}\BibitemShut {NoStop}%
\bibitem [{\citenamefont {Petrovici}(2015)}]{petrovici15}%
  \BibitemOpen
  \bibfield  {author} {\bibinfo {author} {\bibfnamefont {A.}~\bibnamefont
  {Petrovici}},\ }\href {\doibase 10.1103/PhysRevC.91.014302} {\bibfield
  {journal} {\bibinfo  {journal} {Phys. Rev. C}\ }\textbf {\bibinfo {volume}
  {91}},\ \bibinfo {pages} {014302} (\bibinfo {year} {2015})}\BibitemShut
  {NoStop}%
\bibitem [{\citenamefont {Bohr}\ and\ \citenamefont
  {Mottelson}(1975)}]{bohr75}%
  \BibitemOpen
  \bibfield  {author} {\bibinfo {author} {\bibfnamefont {A.}~\bibnamefont
  {Bohr}}\ and\ \bibinfo {author} {\bibfnamefont {B.~R.}\ \bibnamefont
  {Mottelson}},\ }\href@noop {} {\emph {\bibinfo {title} {Nuclear
  Structure}}},\ Vol.~\bibinfo {volume} {2}\ (\bibinfo  {publisher} {Benjamin,
  Reading, Massachusetts},\ \bibinfo {year} {1975})\BibitemShut {NoStop}%
\bibitem [{\citenamefont {Kubo}\ \emph {et~al.}(2012)\citenamefont {Kubo},
  \citenamefont {Kameda}, \citenamefont {Suzuki}, \citenamefont {Fukuda},
  \citenamefont {Takeda}, \citenamefont {Yanagisawa}, \citenamefont {Ohtake},
  \citenamefont {Kusaka}, \citenamefont {Yoshida}, \citenamefont {Inabe},
  \citenamefont {Ohnishi}, \citenamefont {Yoshida}, \citenamefont {Tanaka},\
  and\ \citenamefont {Mizoi}}]{kubo12}%
  \BibitemOpen
  \bibfield  {author} {\bibinfo {author} {\bibfnamefont {T.}~\bibnamefont
  {Kubo}}, \bibinfo {author} {\bibfnamefont {D.}~\bibnamefont {Kameda}},
  \bibinfo {author} {\bibfnamefont {H.}~\bibnamefont {Suzuki}}, \bibinfo
  {author} {\bibfnamefont {N.}~\bibnamefont {Fukuda}}, \bibinfo {author}
  {\bibfnamefont {H.}~\bibnamefont {Takeda}}, \bibinfo {author} {\bibfnamefont
  {Y.}~\bibnamefont {Yanagisawa}}, \bibinfo {author} {\bibfnamefont
  {M.}~\bibnamefont {Ohtake}}, \bibinfo {author} {\bibfnamefont
  {K.}~\bibnamefont {Kusaka}}, \bibinfo {author} {\bibfnamefont
  {K.}~\bibnamefont {Yoshida}}, \bibinfo {author} {\bibfnamefont
  {N.}~\bibnamefont {Inabe}}, \bibinfo {author} {\bibfnamefont
  {T.}~\bibnamefont {Ohnishi}}, \bibinfo {author} {\bibfnamefont
  {A.}~\bibnamefont {Yoshida}}, \bibinfo {author} {\bibfnamefont
  {K.}~\bibnamefont {Tanaka}}, \ and\ \bibinfo {author} {\bibfnamefont
  {Y.}~\bibnamefont {Mizoi}},\ }\href {\doibase 10.1093/ptep/pts064} {\bibfield
   {journal} {\bibinfo  {journal} {Prog. Theo. Exp. Phys.}\ }\textbf {\bibinfo
  {volume} {2012}} (\bibinfo {year} {2012}),\ 10.1093/ptep/pts064},\ \bibinfo
  {note} {03C003},\ \Eprint
  {http://arxiv.org/abs/http://oup.prod.sis.lan/ptep/article-pdf/2012/1/03C003/11595011/pts064.pdf}
  {http://oup.prod.sis.lan/ptep/article-pdf/2012/1/03C003/11595011/pts064.pdf}
  \BibitemShut {NoStop}%
\bibitem [{\citenamefont {Takeuchi}\ \emph {et~al.}(2014)\citenamefont
  {Takeuchi}, \citenamefont {Motobayashi}, \citenamefont {Togano},
  \citenamefont {Matsushita}, \citenamefont {Aoi}, \citenamefont {Demichi},
  \citenamefont {Hasegawa},\ and\ \citenamefont {Murakami}}]{takeuchi14}%
  \BibitemOpen
  \bibfield  {author} {\bibinfo {author} {\bibfnamefont {S.}~\bibnamefont
  {Takeuchi}}, \bibinfo {author} {\bibfnamefont {T.}~\bibnamefont
  {Motobayashi}}, \bibinfo {author} {\bibfnamefont {Y.}~\bibnamefont {Togano}},
  \bibinfo {author} {\bibfnamefont {M.}~\bibnamefont {Matsushita}}, \bibinfo
  {author} {\bibfnamefont {N.}~\bibnamefont {Aoi}}, \bibinfo {author}
  {\bibfnamefont {K.}~\bibnamefont {Demichi}}, \bibinfo {author} {\bibfnamefont
  {H.}~\bibnamefont {Hasegawa}}, \ and\ \bibinfo {author} {\bibfnamefont
  {H.}~\bibnamefont {Murakami}},\ }\href {\doibase
  https://doi.org/10.1016/j.nima.2014.06.087} {\bibfield  {journal} {\bibinfo
  {journal} {Nucl. Instr. Meth. A}\ }\textbf {\bibinfo {volume} {763}},\
  \bibinfo {pages} {596} (\bibinfo {year} {2014})}\BibitemShut {NoStop}%
\bibitem [{\citenamefont {Wimmer}\ \emph {et~al.}(2020)\citenamefont {Wimmer},
  \citenamefont {Arici}, \citenamefont {Korten}, \citenamefont {Doornenbal},
  \citenamefont {Delaroche}, \citenamefont {Girod}, \citenamefont {Libert},
  \citenamefont {Rodr\'iguez}, \citenamefont {Aguilera}, \citenamefont
  {Algora}, \citenamefont {Ando}, \citenamefont {Baba}, \citenamefont {Blank},
  \citenamefont {Boso}, \citenamefont {Chen}, \citenamefont {Corsi},
  \citenamefont {Davies}, \citenamefont {de~Angelis}, \citenamefont
  {de~France}, \citenamefont {Doherty}, \citenamefont {Gerl}, \citenamefont
  {Gernh\"auser}, \citenamefont {Goigoux}, \citenamefont {Jenkins},
  \citenamefont {Kiss}, \citenamefont {Koyama}, \citenamefont {Motobayashi},
  \citenamefont {Nagamine}, \citenamefont {Niikura}, \citenamefont {Nishimura},
  \citenamefont {Obertelli}, \citenamefont {Lubos}, \citenamefont {Phong},
  \citenamefont {Rubio}, \citenamefont {Sahin}, \citenamefont {Saito},
  \citenamefont {Sakurai}, \citenamefont {Sinclair}, \citenamefont
  {Steppenbeck}, \citenamefont {Taniuchi}, \citenamefont {Vaquero},
  \citenamefont {Wadsworth}, \citenamefont {Wu},\ and\ \citenamefont
  {Zielinska}}]{wimmer20}%
  \BibitemOpen
  \bibfield  {author} {\bibinfo {author} {\bibfnamefont {K.}~\bibnamefont
  {Wimmer}}, \bibinfo {author} {\bibfnamefont {T.}~\bibnamefont {Arici}},
  \bibinfo {author} {\bibfnamefont {W.}~\bibnamefont {Korten}}, \bibinfo
  {author} {\bibfnamefont {P.}~\bibnamefont {Doornenbal}}, \bibinfo {author}
  {\bibfnamefont {J.~P.}\ \bibnamefont {Delaroche}}, \bibinfo {author}
  {\bibfnamefont {M.}~\bibnamefont {Girod}}, \bibinfo {author} {\bibfnamefont
  {J.}~\bibnamefont {Libert}}, \bibinfo {author} {\bibfnamefont {T.~R.}\
  \bibnamefont {Rodr\'iguez}}, \bibinfo {author} {\bibfnamefont
  {P.}~\bibnamefont {Aguilera}}, \bibinfo {author} {\bibfnamefont
  {A.}~\bibnamefont {Algora}}, \bibinfo {author} {\bibfnamefont
  {T.}~\bibnamefont {Ando}}, \bibinfo {author} {\bibfnamefont {H.}~\bibnamefont
  {Baba}}, \bibinfo {author} {\bibfnamefont {B.}~\bibnamefont {Blank}},
  \bibinfo {author} {\bibfnamefont {A.}~\bibnamefont {Boso}}, \bibinfo {author}
  {\bibfnamefont {S.}~\bibnamefont {Chen}}, \bibinfo {author} {\bibfnamefont
  {A.}~\bibnamefont {Corsi}}, \bibinfo {author} {\bibfnamefont
  {P.}~\bibnamefont {Davies}}, \bibinfo {author} {\bibfnamefont
  {G.}~\bibnamefont {de~Angelis}}, \bibinfo {author} {\bibfnamefont
  {G.}~\bibnamefont {de~France}}, \bibinfo {author} {\bibfnamefont {D.~T.}\
  \bibnamefont {Doherty}}, \bibinfo {author} {\bibfnamefont {J.}~\bibnamefont
  {Gerl}}, \bibinfo {author} {\bibfnamefont {R.}~\bibnamefont {Gernh\"auser}},
  \bibinfo {author} {\bibfnamefont {T.}~\bibnamefont {Goigoux}}, \bibinfo
  {author} {\bibfnamefont {D.}~\bibnamefont {Jenkins}}, \bibinfo {author}
  {\bibfnamefont {G.}~\bibnamefont {Kiss}}, \bibinfo {author} {\bibfnamefont
  {S.}~\bibnamefont {Koyama}}, \bibinfo {author} {\bibfnamefont
  {T.}~\bibnamefont {Motobayashi}}, \bibinfo {author} {\bibfnamefont
  {S.}~\bibnamefont {Nagamine}}, \bibinfo {author} {\bibfnamefont
  {M.}~\bibnamefont {Niikura}}, \bibinfo {author} {\bibfnamefont
  {S.}~\bibnamefont {Nishimura}}, \bibinfo {author} {\bibfnamefont
  {A.}~\bibnamefont {Obertelli}}, \bibinfo {author} {\bibfnamefont
  {D.}~\bibnamefont {Lubos}}, \bibinfo {author} {\bibfnamefont {V.~H.}\
  \bibnamefont {Phong}}, \bibinfo {author} {\bibfnamefont {B.}~\bibnamefont
  {Rubio}}, \bibinfo {author} {\bibfnamefont {E.}~\bibnamefont {Sahin}},
  \bibinfo {author} {\bibfnamefont {T.~Y.}\ \bibnamefont {Saito}}, \bibinfo
  {author} {\bibfnamefont {H.}~\bibnamefont {Sakurai}}, \bibinfo {author}
  {\bibfnamefont {L.}~\bibnamefont {Sinclair}}, \bibinfo {author}
  {\bibfnamefont {D.}~\bibnamefont {Steppenbeck}}, \bibinfo {author}
  {\bibfnamefont {R.}~\bibnamefont {Taniuchi}}, \bibinfo {author}
  {\bibfnamefont {V.}~\bibnamefont {Vaquero}}, \bibinfo {author} {\bibfnamefont
  {R.}~\bibnamefont {Wadsworth}}, \bibinfo {author} {\bibfnamefont
  {J.}~\bibnamefont {Wu}}, \ and\ \bibinfo {author} {\bibfnamefont
  {M.}~\bibnamefont {Zielinska}},\ }\href {\doibase
  10.1140/epja/s10050-020-00171-3} {\bibfield  {journal} {\bibinfo  {journal}
  {Eur. Phys. Jour. A}\ }\textbf {\bibinfo {volume} {56}},\ \bibinfo {pages}
  {159} (\bibinfo {year} {2020})}\BibitemShut {NoStop}%
\bibitem [{\citenamefont {Karny}\ \emph {et~al.}(2004)\citenamefont {Karny},
  \citenamefont {Batist}, \citenamefont {Jenkins}, \citenamefont {Kavatsyuk},
  \citenamefont {Kavatsyuk}, \citenamefont {Kirchner}, \citenamefont {Korgul},
  \citenamefont {Roeckl},\ and\ \citenamefont {\ifmmode~\dot{Z}\else
  \.{Z}\fi{}ylicz}}]{karny04}%
  \BibitemOpen
  \bibfield  {author} {\bibinfo {author} {\bibfnamefont {M.}~\bibnamefont
  {Karny}}, \bibinfo {author} {\bibfnamefont {L.}~\bibnamefont {Batist}},
  \bibinfo {author} {\bibfnamefont {D.}~\bibnamefont {Jenkins}}, \bibinfo
  {author} {\bibfnamefont {M.}~\bibnamefont {Kavatsyuk}}, \bibinfo {author}
  {\bibfnamefont {O.}~\bibnamefont {Kavatsyuk}}, \bibinfo {author}
  {\bibfnamefont {R.}~\bibnamefont {Kirchner}}, \bibinfo {author}
  {\bibfnamefont {A.}~\bibnamefont {Korgul}}, \bibinfo {author} {\bibfnamefont
  {E.}~\bibnamefont {Roeckl}}, \ and\ \bibinfo {author} {\bibfnamefont
  {J.}~\bibnamefont {\ifmmode~\dot{Z}\else \.{Z}\fi{}ylicz}},\ }\href {\doibase
  10.1103/PhysRevC.70.014310} {\bibfield  {journal} {\bibinfo  {journal} {Phys.
  Rev. C}\ }\textbf {\bibinfo {volume} {70}},\ \bibinfo {pages} {014310}
  (\bibinfo {year} {2004})}\BibitemShut {NoStop}%
\bibitem [{\citenamefont {Wadsworth}\ \emph {et~al.}(1980)\citenamefont
  {Wadsworth}, \citenamefont {Ekstrom}, \citenamefont {Jones}, \citenamefont
  {Kearns}, \citenamefont {Morrison}, \citenamefont {Twin},\ and\ \citenamefont
  {Ward}}]{wadsworth80}%
  \BibitemOpen
  \bibfield  {author} {\bibinfo {author} {\bibfnamefont {R.}~\bibnamefont
  {Wadsworth}}, \bibinfo {author} {\bibfnamefont {L.~P.}\ \bibnamefont
  {Ekstrom}}, \bibinfo {author} {\bibfnamefont {G.~D.}\ \bibnamefont {Jones}},
  \bibinfo {author} {\bibfnamefont {F.}~\bibnamefont {Kearns}}, \bibinfo
  {author} {\bibfnamefont {T.~P.}\ \bibnamefont {Morrison}}, \bibinfo {author}
  {\bibfnamefont {P.~J.}\ \bibnamefont {Twin}}, \ and\ \bibinfo {author}
  {\bibfnamefont {N.~J.}\ \bibnamefont {Ward}},\ }\href {\doibase
  10.1088/0305-4616/6/11/012} {\bibfield  {journal} {\bibinfo  {journal} {J.
  Phys. G}\ ,\ \bibinfo {pages} {1403}} (\bibinfo {year} {1980})}\BibitemShut
  {NoStop}%
\bibitem [{\citenamefont {Delaroche}\ \emph {et~al.}(2010)\citenamefont
  {Delaroche}, \citenamefont {Girod}, \citenamefont {Libert}, \citenamefont
  {Goutte}, \citenamefont {Hilaire}, \citenamefont {P\'eru}, \citenamefont
  {Pillet},\ and\ \citenamefont {Bertsch}}]{delaroche10}%
  \BibitemOpen
  \bibfield  {author} {\bibinfo {author} {\bibfnamefont {J.~P.}\ \bibnamefont
  {Delaroche}}, \bibinfo {author} {\bibfnamefont {M.}~\bibnamefont {Girod}},
  \bibinfo {author} {\bibfnamefont {J.}~\bibnamefont {Libert}}, \bibinfo
  {author} {\bibfnamefont {H.}~\bibnamefont {Goutte}}, \bibinfo {author}
  {\bibfnamefont {S.}~\bibnamefont {Hilaire}}, \bibinfo {author} {\bibfnamefont
  {S.}~\bibnamefont {P\'eru}}, \bibinfo {author} {\bibfnamefont
  {N.}~\bibnamefont {Pillet}}, \ and\ \bibinfo {author} {\bibfnamefont {G.~F.}\
  \bibnamefont {Bertsch}},\ }\href {\doibase 10.1103/PhysRevC.81.014303}
  {\bibfield  {journal} {\bibinfo  {journal} {Phys. Rev. C}\ }\textbf {\bibinfo
  {volume} {81}},\ \bibinfo {pages} {014303} (\bibinfo {year}
  {2010})}\BibitemShut {NoStop}%
\bibitem [{\citenamefont {Rodr\'{\i}guez}(2014)}]{rodriguez14}%
  \BibitemOpen
  \bibfield  {author} {\bibinfo {author} {\bibfnamefont {T.~R.}\ \bibnamefont
  {Rodr\'{\i}guez}},\ }\href@noop {} {\bibfield  {journal} {\bibinfo  {journal}
  {Phys. Rev. C}\ }\textbf {\bibinfo {volume} {90}},\ \bibinfo {pages} {034306}
  (\bibinfo {year} {2014})}\BibitemShut {NoStop}%
\bibitem [{\citenamefont {Thompson}(1988)}]{thompson88}%
  \BibitemOpen
  \bibfield  {author} {\bibinfo {author} {\bibfnamefont {I.~J.}\ \bibnamefont
  {Thompson}},\ }\href
  {http://www.sciencedirect.com/science/article/pii/0167797788900056}
  {\bibfield  {journal} {\bibinfo  {journal} {Comp. Phys. Rep.}\ }\textbf
  {\bibinfo {volume} {7}},\ \bibinfo {pages} {167} (\bibinfo {year}
  {1988})}\BibitemShut {NoStop}%
\bibitem [{\citenamefont {Moro}(2018)}]{moro18}%
  \BibitemOpen
  \bibfield  {author} {\bibinfo {author} {\bibfnamefont {A.}~\bibnamefont
  {Moro}},\ }\href@noop {} {} (\bibinfo {year} {2018}),\ \bibinfo {note} {priv.
  comm.}\BibitemShut {Stop}%
\bibitem [{\citenamefont {Furumoto}\ \emph {et~al.}(2012)\citenamefont
  {Furumoto}, \citenamefont {Horiuchi}, \citenamefont {Takashina},
  \citenamefont {Yamamoto},\ and\ \citenamefont {Sakuragi}}]{furumoto12}%
  \BibitemOpen
  \bibfield  {author} {\bibinfo {author} {\bibfnamefont {T.}~\bibnamefont
  {Furumoto}}, \bibinfo {author} {\bibfnamefont {W.}~\bibnamefont {Horiuchi}},
  \bibinfo {author} {\bibfnamefont {M.}~\bibnamefont {Takashina}}, \bibinfo
  {author} {\bibfnamefont {Y.}~\bibnamefont {Yamamoto}}, \ and\ \bibinfo
  {author} {\bibfnamefont {Y.}~\bibnamefont {Sakuragi}},\ }\href {\doibase
  10.1103/PhysRevC.85.044607} {\bibfield  {journal} {\bibinfo  {journal} {Phys.
  Rev. C}\ }\textbf {\bibinfo {volume} {85}},\ \bibinfo {pages} {044607}
  (\bibinfo {year} {2012})}\BibitemShut {NoStop}%
\bibitem [{\citenamefont {Gade}\ \emph {et~al.}(2005)\citenamefont {Gade},
  \citenamefont {Bazin}, \citenamefont {Becerril}, \citenamefont {Campbell},
  \citenamefont {Cook}, \citenamefont {Dean}, \citenamefont {Dinca},
  \citenamefont {Glasmacher}, \citenamefont {Hitt}, \citenamefont {Howard},
  \citenamefont {Mueller}, \citenamefont {Olliver}, \citenamefont {Terry},\
  and\ \citenamefont {Yoneda}}]{gade05}%
  \BibitemOpen
  \bibfield  {author} {\bibinfo {author} {\bibfnamefont {A.}~\bibnamefont
  {Gade}}, \bibinfo {author} {\bibfnamefont {D.}~\bibnamefont {Bazin}},
  \bibinfo {author} {\bibfnamefont {A.}~\bibnamefont {Becerril}}, \bibinfo
  {author} {\bibfnamefont {C.~M.}\ \bibnamefont {Campbell}}, \bibinfo {author}
  {\bibfnamefont {J.~M.}\ \bibnamefont {Cook}}, \bibinfo {author}
  {\bibfnamefont {D.~J.}\ \bibnamefont {Dean}}, \bibinfo {author}
  {\bibfnamefont {D.-C.}\ \bibnamefont {Dinca}}, \bibinfo {author}
  {\bibfnamefont {T.}~\bibnamefont {Glasmacher}}, \bibinfo {author}
  {\bibfnamefont {G.~W.}\ \bibnamefont {Hitt}}, \bibinfo {author}
  {\bibfnamefont {M.~E.}\ \bibnamefont {Howard}}, \bibinfo {author}
  {\bibfnamefont {W.~F.}\ \bibnamefont {Mueller}}, \bibinfo {author}
  {\bibfnamefont {H.}~\bibnamefont {Olliver}}, \bibinfo {author} {\bibfnamefont
  {J.~R.}\ \bibnamefont {Terry}}, \ and\ \bibinfo {author} {\bibfnamefont
  {K.}~\bibnamefont {Yoneda}},\ }\href {\doibase 10.1103/PhysRevLett.95.022502}
  {\bibfield  {journal} {\bibinfo  {journal} {Phys. Rev. Lett.}\ }\textbf
  {\bibinfo {volume} {95}},\ \bibinfo {pages} {022502} (\bibinfo {year}
  {2005})}\BibitemShut {NoStop}%
\bibitem [{\citenamefont {Iwasaki}\ \emph {et~al.}(2014)\citenamefont
  {Iwasaki}, \citenamefont {Lemasson}, \citenamefont {Morse}, \citenamefont
  {Dewald}, \citenamefont {Braunroth}, \citenamefont {Bader}, \citenamefont
  {Baugher}, \citenamefont {Bazin}, \citenamefont {Berryman}, \citenamefont
  {Campbell}, \citenamefont {Gade}, \citenamefont {Langer}, \citenamefont
  {Lee}, \citenamefont {Loelius}, \citenamefont {Lunderberg}, \citenamefont
  {Recchia}, \citenamefont {Smalley}, \citenamefont {Stroberg}, \citenamefont
  {Wadsworth}, \citenamefont {Walz}, \citenamefont {Weisshaar}, \citenamefont
  {Westerberg}, \citenamefont {Whitmore},\ and\ \citenamefont
  {Wimmer}}]{iwasaki14}%
  \BibitemOpen
  \bibfield  {author} {\bibinfo {author} {\bibfnamefont {H.}~\bibnamefont
  {Iwasaki}}, \bibinfo {author} {\bibfnamefont {A.}~\bibnamefont {Lemasson}},
  \bibinfo {author} {\bibfnamefont {C.}~\bibnamefont {Morse}}, \bibinfo
  {author} {\bibfnamefont {A.}~\bibnamefont {Dewald}}, \bibinfo {author}
  {\bibfnamefont {T.}~\bibnamefont {Braunroth}}, \bibinfo {author}
  {\bibfnamefont {V.~M.}\ \bibnamefont {Bader}}, \bibinfo {author}
  {\bibfnamefont {T.}~\bibnamefont {Baugher}}, \bibinfo {author} {\bibfnamefont
  {D.}~\bibnamefont {Bazin}}, \bibinfo {author} {\bibfnamefont {J.~S.}\
  \bibnamefont {Berryman}}, \bibinfo {author} {\bibfnamefont {C.~M.}\
  \bibnamefont {Campbell}}, \bibinfo {author} {\bibfnamefont {A.}~\bibnamefont
  {Gade}}, \bibinfo {author} {\bibfnamefont {C.}~\bibnamefont {Langer}},
  \bibinfo {author} {\bibfnamefont {I.~Y.}\ \bibnamefont {Lee}}, \bibinfo
  {author} {\bibfnamefont {C.}~\bibnamefont {Loelius}}, \bibinfo {author}
  {\bibfnamefont {E.}~\bibnamefont {Lunderberg}}, \bibinfo {author}
  {\bibfnamefont {F.}~\bibnamefont {Recchia}}, \bibinfo {author} {\bibfnamefont
  {D.}~\bibnamefont {Smalley}}, \bibinfo {author} {\bibfnamefont {S.~R.}\
  \bibnamefont {Stroberg}}, \bibinfo {author} {\bibfnamefont {R.}~\bibnamefont
  {Wadsworth}}, \bibinfo {author} {\bibfnamefont {C.}~\bibnamefont {Walz}},
  \bibinfo {author} {\bibfnamefont {D.}~\bibnamefont {Weisshaar}}, \bibinfo
  {author} {\bibfnamefont {A.}~\bibnamefont {Westerberg}}, \bibinfo {author}
  {\bibfnamefont {K.}~\bibnamefont {Whitmore}}, \ and\ \bibinfo {author}
  {\bibfnamefont {K.}~\bibnamefont {Wimmer}},\ }\href {\doibase
  10.1103/PhysRevLett.112.142502} {\bibfield  {journal} {\bibinfo  {journal}
  {Phys. Rev. Lett.}\ }\textbf {\bibinfo {volume} {112}},\ \bibinfo {pages}
  {142502} (\bibinfo {year} {2014})}\BibitemShut {NoStop}%
\bibitem [{\citenamefont {Obertelli}\ \emph {et~al.}(2009)\citenamefont
  {Obertelli}, \citenamefont {Baugher}, \citenamefont {Bazin}, \citenamefont
  {Delaroche}, \citenamefont {Flavigny}, \citenamefont {Gade}, \citenamefont
  {Girod}, \citenamefont {Glasmacher}, \citenamefont {Goergen}, \citenamefont
  {Grinyer}, \citenamefont {Korten}, \citenamefont {Ljungvall}, \citenamefont
  {McDaniel}, \citenamefont {Ratkiewicz}, \citenamefont {Sulignano},\ and\
  \citenamefont {Weisshaar}}]{obertelli09}%
  \BibitemOpen
  \bibfield  {author} {\bibinfo {author} {\bibfnamefont {A.}~\bibnamefont
  {Obertelli}}, \bibinfo {author} {\bibfnamefont {T.}~\bibnamefont {Baugher}},
  \bibinfo {author} {\bibfnamefont {D.}~\bibnamefont {Bazin}}, \bibinfo
  {author} {\bibfnamefont {J.~P.}\ \bibnamefont {Delaroche}}, \bibinfo {author}
  {\bibfnamefont {F.}~\bibnamefont {Flavigny}}, \bibinfo {author}
  {\bibfnamefont {A.}~\bibnamefont {Gade}}, \bibinfo {author} {\bibfnamefont
  {M.}~\bibnamefont {Girod}}, \bibinfo {author} {\bibfnamefont
  {T.}~\bibnamefont {Glasmacher}}, \bibinfo {author} {\bibfnamefont
  {A.}~\bibnamefont {Goergen}}, \bibinfo {author} {\bibfnamefont {G.~F.}\
  \bibnamefont {Grinyer}}, \bibinfo {author} {\bibfnamefont {W.}~\bibnamefont
  {Korten}}, \bibinfo {author} {\bibfnamefont {J.}~\bibnamefont {Ljungvall}},
  \bibinfo {author} {\bibfnamefont {S.}~\bibnamefont {McDaniel}}, \bibinfo
  {author} {\bibfnamefont {A.}~\bibnamefont {Ratkiewicz}}, \bibinfo {author}
  {\bibfnamefont {B.}~\bibnamefont {Sulignano}}, \ and\ \bibinfo {author}
  {\bibfnamefont {D.}~\bibnamefont {Weisshaar}},\ }\href {\doibase
  10.1103/PhysRevC.80.031304} {\bibfield  {journal} {\bibinfo  {journal} {Phys.
  Rev. C}\ }\textbf {\bibinfo {volume} {80}},\ \bibinfo {pages} {031304}
  (\bibinfo {year} {2009})}\BibitemShut {NoStop}%
\bibitem [{\citenamefont {Nichols}\ \emph {et~al.}(2014)\citenamefont
  {Nichols}, \citenamefont {Wadsworth}, \citenamefont {Iwasaki}, \citenamefont
  {Kaneko}, \citenamefont {Lemasson}, \citenamefont {de~Angelis}, \citenamefont
  {Bader}, \citenamefont {Baugher}, \citenamefont {Bazin}, \citenamefont
  {Bentley}, \citenamefont {Berryman}, \citenamefont {Braunroth}, \citenamefont
  {Davies}, \citenamefont {Dewald}, \citenamefont {Fransen}, \citenamefont
  {Gade}, \citenamefont {Hackstein}, \citenamefont {Henderson}, \citenamefont
  {Jenkins}, \citenamefont {Miller}, \citenamefont {Morse}, \citenamefont
  {Paterson}, \citenamefont {Simpson}, \citenamefont {Stroberg}, \citenamefont
  {Weisshaar}, \citenamefont {Whitmore},\ and\ \citenamefont
  {Wimmer}}]{nichols14}%
  \BibitemOpen
  \bibfield  {author} {\bibinfo {author} {\bibfnamefont {A.}~\bibnamefont
  {Nichols}}, \bibinfo {author} {\bibfnamefont {R.}~\bibnamefont {Wadsworth}},
  \bibinfo {author} {\bibfnamefont {H.}~\bibnamefont {Iwasaki}}, \bibinfo
  {author} {\bibfnamefont {K.}~\bibnamefont {Kaneko}}, \bibinfo {author}
  {\bibfnamefont {A.}~\bibnamefont {Lemasson}}, \bibinfo {author}
  {\bibfnamefont {G.}~\bibnamefont {de~Angelis}}, \bibinfo {author}
  {\bibfnamefont {V.}~\bibnamefont {Bader}}, \bibinfo {author} {\bibfnamefont
  {T.}~\bibnamefont {Baugher}}, \bibinfo {author} {\bibfnamefont
  {D.}~\bibnamefont {Bazin}}, \bibinfo {author} {\bibfnamefont
  {M.}~\bibnamefont {Bentley}}, \bibinfo {author} {\bibfnamefont
  {J.}~\bibnamefont {Berryman}}, \bibinfo {author} {\bibfnamefont
  {T.}~\bibnamefont {Braunroth}}, \bibinfo {author} {\bibfnamefont
  {P.}~\bibnamefont {Davies}}, \bibinfo {author} {\bibfnamefont
  {A.}~\bibnamefont {Dewald}}, \bibinfo {author} {\bibfnamefont
  {C.}~\bibnamefont {Fransen}}, \bibinfo {author} {\bibfnamefont
  {A.}~\bibnamefont {Gade}}, \bibinfo {author} {\bibfnamefont {M.}~\bibnamefont
  {Hackstein}}, \bibinfo {author} {\bibfnamefont {J.}~\bibnamefont
  {Henderson}}, \bibinfo {author} {\bibfnamefont {D.}~\bibnamefont {Jenkins}},
  \bibinfo {author} {\bibfnamefont {D.}~\bibnamefont {Miller}}, \bibinfo
  {author} {\bibfnamefont {C.}~\bibnamefont {Morse}}, \bibinfo {author}
  {\bibfnamefont {I.}~\bibnamefont {Paterson}}, \bibinfo {author}
  {\bibfnamefont {E.}~\bibnamefont {Simpson}}, \bibinfo {author} {\bibfnamefont
  {S.}~\bibnamefont {Stroberg}}, \bibinfo {author} {\bibfnamefont
  {D.}~\bibnamefont {Weisshaar}}, \bibinfo {author} {\bibfnamefont
  {K.}~\bibnamefont {Whitmore}}, \ and\ \bibinfo {author} {\bibfnamefont
  {K.}~\bibnamefont {Wimmer}},\ }\href {\doibase
  https://doi.org/10.1016/j.physletb.2014.04.016} {\bibfield  {journal}
  {\bibinfo  {journal} {Phys. Lett. B}\ }\textbf {\bibinfo {volume} {733}},\
  \bibinfo {pages} {52} (\bibinfo {year} {2014})}\BibitemShut {NoStop}%
\bibitem [{\citenamefont {Ljungvall}\ \emph {et~al.}(2008)\citenamefont
  {Ljungvall}, \citenamefont {G\"orgen}, \citenamefont {Girod}, \citenamefont
  {Delaroche}, \citenamefont {Dewald}, \citenamefont {Dossat}, \citenamefont
  {Farnea}, \citenamefont {Korten}, \citenamefont {Melon}, \citenamefont
  {Menegazzo}, \citenamefont {Obertelli}, \citenamefont {Orlandi},
  \citenamefont {Petkov}, \citenamefont {Pissulla}, \citenamefont {Siem},
  \citenamefont {Singh}, \citenamefont {Srebrny}, \citenamefont {Theisen},
  \citenamefont {Ur}, \citenamefont {Valiente-Dob\'on}, \citenamefont {Zell},\
  and\ \citenamefont {Zieli\ifmmode~\acute{n}\else
  \'{n}\fi{}ska}}]{ljungvall08}%
  \BibitemOpen
  \bibfield  {author} {\bibinfo {author} {\bibfnamefont {J.}~\bibnamefont
  {Ljungvall}}, \bibinfo {author} {\bibfnamefont {A.}~\bibnamefont {G\"orgen}},
  \bibinfo {author} {\bibfnamefont {M.}~\bibnamefont {Girod}}, \bibinfo
  {author} {\bibfnamefont {J.-P.}\ \bibnamefont {Delaroche}}, \bibinfo {author}
  {\bibfnamefont {A.}~\bibnamefont {Dewald}}, \bibinfo {author} {\bibfnamefont
  {C.}~\bibnamefont {Dossat}}, \bibinfo {author} {\bibfnamefont
  {E.}~\bibnamefont {Farnea}}, \bibinfo {author} {\bibfnamefont
  {W.}~\bibnamefont {Korten}}, \bibinfo {author} {\bibfnamefont
  {B.}~\bibnamefont {Melon}}, \bibinfo {author} {\bibfnamefont
  {R.}~\bibnamefont {Menegazzo}}, \bibinfo {author} {\bibfnamefont
  {A.}~\bibnamefont {Obertelli}}, \bibinfo {author} {\bibfnamefont
  {R.}~\bibnamefont {Orlandi}}, \bibinfo {author} {\bibfnamefont
  {P.}~\bibnamefont {Petkov}}, \bibinfo {author} {\bibfnamefont
  {T.}~\bibnamefont {Pissulla}}, \bibinfo {author} {\bibfnamefont
  {S.}~\bibnamefont {Siem}}, \bibinfo {author} {\bibfnamefont {R.~P.}\
  \bibnamefont {Singh}}, \bibinfo {author} {\bibfnamefont {J.}~\bibnamefont
  {Srebrny}}, \bibinfo {author} {\bibfnamefont {C.}~\bibnamefont {Theisen}},
  \bibinfo {author} {\bibfnamefont {C.~A.}\ \bibnamefont {Ur}}, \bibinfo
  {author} {\bibfnamefont {J.~J.}\ \bibnamefont {Valiente-Dob\'on}}, \bibinfo
  {author} {\bibfnamefont {K.~O.}\ \bibnamefont {Zell}}, \ and\ \bibinfo
  {author} {\bibfnamefont {M.}~\bibnamefont {Zieli\ifmmode~\acute{n}\else
  \'{n}\fi{}ska}},\ }\href {\doibase 10.1103/PhysRevLett.100.102502} {\bibfield
   {journal} {\bibinfo  {journal} {Phys. Rev. Lett.}\ }\textbf {\bibinfo
  {volume} {100}},\ \bibinfo {pages} {102502} (\bibinfo {year}
  {2008})}\BibitemShut {NoStop}%
\bibitem [{\citenamefont {Morse}\ \emph {et~al.}(2018)\citenamefont {Morse},
  \citenamefont {Iwasaki}, \citenamefont {Lemasson}, \citenamefont {Dewald},
  \citenamefont {Braunroth}, \citenamefont {Bader}, \citenamefont {Baugher},
  \citenamefont {Bazin}, \citenamefont {Berryman}, \citenamefont {Campbell},
  \citenamefont {Gade}, \citenamefont {Langer}, \citenamefont {Lee},
  \citenamefont {Loelius}, \citenamefont {Lunderberg}, \citenamefont {Recchia},
  \citenamefont {Smalley}, \citenamefont {Stroberg}, \citenamefont {Wadsworth},
  \citenamefont {Walz}, \citenamefont {Weisshaar}, \citenamefont {Westerberg},
  \citenamefont {Whitmore},\ and\ \citenamefont {Wimmer}}]{morse18}%
  \BibitemOpen
  \bibfield  {author} {\bibinfo {author} {\bibfnamefont {C.}~\bibnamefont
  {Morse}}, \bibinfo {author} {\bibfnamefont {H.}~\bibnamefont {Iwasaki}},
  \bibinfo {author} {\bibfnamefont {A.}~\bibnamefont {Lemasson}}, \bibinfo
  {author} {\bibfnamefont {A.}~\bibnamefont {Dewald}}, \bibinfo {author}
  {\bibfnamefont {T.}~\bibnamefont {Braunroth}}, \bibinfo {author}
  {\bibfnamefont {V.}~\bibnamefont {Bader}}, \bibinfo {author} {\bibfnamefont
  {T.}~\bibnamefont {Baugher}}, \bibinfo {author} {\bibfnamefont
  {D.}~\bibnamefont {Bazin}}, \bibinfo {author} {\bibfnamefont
  {J.}~\bibnamefont {Berryman}}, \bibinfo {author} {\bibfnamefont
  {C.}~\bibnamefont {Campbell}}, \bibinfo {author} {\bibfnamefont
  {A.}~\bibnamefont {Gade}}, \bibinfo {author} {\bibfnamefont {C.}~\bibnamefont
  {Langer}}, \bibinfo {author} {\bibfnamefont {I.}~\bibnamefont {Lee}},
  \bibinfo {author} {\bibfnamefont {C.}~\bibnamefont {Loelius}}, \bibinfo
  {author} {\bibfnamefont {E.}~\bibnamefont {Lunderberg}}, \bibinfo {author}
  {\bibfnamefont {F.}~\bibnamefont {Recchia}}, \bibinfo {author} {\bibfnamefont
  {D.}~\bibnamefont {Smalley}}, \bibinfo {author} {\bibfnamefont
  {S.}~\bibnamefont {Stroberg}}, \bibinfo {author} {\bibfnamefont
  {R.}~\bibnamefont {Wadsworth}}, \bibinfo {author} {\bibfnamefont
  {C.}~\bibnamefont {Walz}}, \bibinfo {author} {\bibfnamefont {D.}~\bibnamefont
  {Weisshaar}}, \bibinfo {author} {\bibfnamefont {A.}~\bibnamefont
  {Westerberg}}, \bibinfo {author} {\bibfnamefont {K.}~\bibnamefont
  {Whitmore}}, \ and\ \bibinfo {author} {\bibfnamefont {K.}~\bibnamefont
  {Wimmer}},\ }\href {\doibase https://doi.org/10.1016/j.physletb.2018.10.064}
  {\bibfield  {journal} {\bibinfo  {journal} {Phys. Lett. B}\ }\textbf
  {\bibinfo {volume} {787}},\ \bibinfo {pages} {198} (\bibinfo {year}
  {2018})}\BibitemShut {NoStop}%
\bibitem [{\citenamefont {Lemasson}\ \emph {et~al.}(2012)\citenamefont
  {Lemasson}, \citenamefont {Iwasaki}, \citenamefont {Morse}, \citenamefont
  {Bazin}, \citenamefont {Baugher}, \citenamefont {Berryman}, \citenamefont
  {Dewald}, \citenamefont {Fransen}, \citenamefont {Gade}, \citenamefont
  {McDaniel}, \citenamefont {Nichols}, \citenamefont {Ratkiewicz},
  \citenamefont {Stroberg}, \citenamefont {Voss}, \citenamefont {Wadsworth},
  \citenamefont {Weisshaar}, \citenamefont {Wimmer},\ and\ \citenamefont
  {Winkler}}]{lemasson12}%
  \BibitemOpen
  \bibfield  {author} {\bibinfo {author} {\bibfnamefont {A.}~\bibnamefont
  {Lemasson}}, \bibinfo {author} {\bibfnamefont {H.}~\bibnamefont {Iwasaki}},
  \bibinfo {author} {\bibfnamefont {C.}~\bibnamefont {Morse}}, \bibinfo
  {author} {\bibfnamefont {D.}~\bibnamefont {Bazin}}, \bibinfo {author}
  {\bibfnamefont {T.}~\bibnamefont {Baugher}}, \bibinfo {author} {\bibfnamefont
  {J.~S.}\ \bibnamefont {Berryman}}, \bibinfo {author} {\bibfnamefont
  {A.}~\bibnamefont {Dewald}}, \bibinfo {author} {\bibfnamefont
  {C.}~\bibnamefont {Fransen}}, \bibinfo {author} {\bibfnamefont
  {A.}~\bibnamefont {Gade}}, \bibinfo {author} {\bibfnamefont {S.}~\bibnamefont
  {McDaniel}}, \bibinfo {author} {\bibfnamefont {A.}~\bibnamefont {Nichols}},
  \bibinfo {author} {\bibfnamefont {A.}~\bibnamefont {Ratkiewicz}}, \bibinfo
  {author} {\bibfnamefont {S.}~\bibnamefont {Stroberg}}, \bibinfo {author}
  {\bibfnamefont {P.}~\bibnamefont {Voss}}, \bibinfo {author} {\bibfnamefont
  {R.}~\bibnamefont {Wadsworth}}, \bibinfo {author} {\bibfnamefont
  {D.}~\bibnamefont {Weisshaar}}, \bibinfo {author} {\bibfnamefont
  {K.}~\bibnamefont {Wimmer}}, \ and\ \bibinfo {author} {\bibfnamefont
  {R.}~\bibnamefont {Winkler}},\ }\href {\doibase 10.1103/PhysRevC.85.041303}
  {\bibfield  {journal} {\bibinfo  {journal} {Phys. Rev. C}\ }\textbf {\bibinfo
  {volume} {85}},\ \bibinfo {pages} {041303} (\bibinfo {year}
  {2012})}\BibitemShut {NoStop}%
\bibitem [{\citenamefont {Llewellyn}\ \emph {et~al.}(2020)\citenamefont
  {Llewellyn}, \citenamefont {Bentley}, \citenamefont {Wadsworth},
  \citenamefont {Iwasaki}, \citenamefont {Dobaczewski}, \citenamefont
  {de~Angelis}, \citenamefont {Ash}, \citenamefont {Bazin}, \citenamefont
  {Bender}, \citenamefont {Cederwall}, \citenamefont {Crider}, \citenamefont
  {Doncel}, \citenamefont {Elder}, \citenamefont {Elman}, \citenamefont {Gade},
  \citenamefont {Grinder}, \citenamefont {Haylett}, \citenamefont {Jenkins},
  \citenamefont {Lee}, \citenamefont {Longfellow}, \citenamefont {Lunderberg},
  \citenamefont {Mijatovi\ifmmode~\acute{c}\else \'{c}\fi{}}, \citenamefont
  {Milne}, \citenamefont {Muir}, \citenamefont {Pastore}, \citenamefont
  {Rhodes},\ and\ \citenamefont {Weisshaar}}]{llewellyn20}%
  \BibitemOpen
  \bibfield  {author} {\bibinfo {author} {\bibfnamefont {R.~D.~O.}\
  \bibnamefont {Llewellyn}}, \bibinfo {author} {\bibfnamefont {M.~A.}\
  \bibnamefont {Bentley}}, \bibinfo {author} {\bibfnamefont {R.}~\bibnamefont
  {Wadsworth}}, \bibinfo {author} {\bibfnamefont {H.}~\bibnamefont {Iwasaki}},
  \bibinfo {author} {\bibfnamefont {J.}~\bibnamefont {Dobaczewski}}, \bibinfo
  {author} {\bibfnamefont {G.}~\bibnamefont {de~Angelis}}, \bibinfo {author}
  {\bibfnamefont {J.}~\bibnamefont {Ash}}, \bibinfo {author} {\bibfnamefont
  {D.}~\bibnamefont {Bazin}}, \bibinfo {author} {\bibfnamefont {P.~C.}\
  \bibnamefont {Bender}}, \bibinfo {author} {\bibfnamefont {B.}~\bibnamefont
  {Cederwall}}, \bibinfo {author} {\bibfnamefont {B.~P.}\ \bibnamefont
  {Crider}}, \bibinfo {author} {\bibfnamefont {M.}~\bibnamefont {Doncel}},
  \bibinfo {author} {\bibfnamefont {R.}~\bibnamefont {Elder}}, \bibinfo
  {author} {\bibfnamefont {B.}~\bibnamefont {Elman}}, \bibinfo {author}
  {\bibfnamefont {A.}~\bibnamefont {Gade}}, \bibinfo {author} {\bibfnamefont
  {M.}~\bibnamefont {Grinder}}, \bibinfo {author} {\bibfnamefont
  {T.}~\bibnamefont {Haylett}}, \bibinfo {author} {\bibfnamefont {D.~G.}\
  \bibnamefont {Jenkins}}, \bibinfo {author} {\bibfnamefont {I.~Y.}\
  \bibnamefont {Lee}}, \bibinfo {author} {\bibfnamefont {B.}~\bibnamefont
  {Longfellow}}, \bibinfo {author} {\bibfnamefont {E.}~\bibnamefont
  {Lunderberg}}, \bibinfo {author} {\bibfnamefont {T.}~\bibnamefont
  {Mijatovi\ifmmode~\acute{c}\else \'{c}\fi{}}}, \bibinfo {author}
  {\bibfnamefont {S.~A.}\ \bibnamefont {Milne}}, \bibinfo {author}
  {\bibfnamefont {D.}~\bibnamefont {Muir}}, \bibinfo {author} {\bibfnamefont
  {A.}~\bibnamefont {Pastore}}, \bibinfo {author} {\bibfnamefont
  {D.}~\bibnamefont {Rhodes}}, \ and\ \bibinfo {author} {\bibfnamefont
  {D.}~\bibnamefont {Weisshaar}},\ }\href@noop {} {\bibfield  {journal}
  {\bibinfo  {journal} {Phys. Rev. Lett.}\ }\textbf {\bibinfo {volume} {124}},\
  \bibinfo {pages} {152501} (\bibinfo {year} {2020})}\BibitemShut {NoStop}%
\bibitem [{\citenamefont {M{\"o}ller}\ \emph {et~al.}(2016)\citenamefont
  {M{\"o}ller}, \citenamefont {Sierk}, \citenamefont {Ichikawa},\ and\
  \citenamefont {Sagawa}}]{moeller16}%
  \BibitemOpen
  \bibfield  {author} {\bibinfo {author} {\bibfnamefont {P.}~\bibnamefont
  {M{\"o}ller}}, \bibinfo {author} {\bibfnamefont {A.}~\bibnamefont {Sierk}},
  \bibinfo {author} {\bibfnamefont {T.}~\bibnamefont {Ichikawa}}, \ and\
  \bibinfo {author} {\bibfnamefont {H.}~\bibnamefont {Sagawa}},\ }\href
  {http://www.sciencedirect.com/science/article/pii/S0092640X1600005X}
  {\bibfield  {journal} {\bibinfo  {journal} {At. Data Nucl. Data Tables}\
  }\textbf {\bibinfo {volume} {109}},\ \bibinfo {pages} {1} (\bibinfo {year}
  {2016})}\BibitemShut {NoStop}%
\bibitem [{\citenamefont {Honma}\ \emph {et~al.}(2005)\citenamefont {Honma},
  \citenamefont {Otsuka}, \citenamefont {Brown},\ and\ \citenamefont
  {Mizusaki}}]{honma05}%
  \BibitemOpen
  \bibfield  {author} {\bibinfo {author} {\bibfnamefont {M.}~\bibnamefont
  {Honma}}, \bibinfo {author} {\bibfnamefont {T.}~\bibnamefont {Otsuka}},
  \bibinfo {author} {\bibfnamefont {B.~A.}\ \bibnamefont {Brown}}, \ and\
  \bibinfo {author} {\bibfnamefont {T.}~\bibnamefont {Mizusaki}},\ }\href
  {\doibase https://doi.org/10.1140/epjad/i2005-06-032-2} {\bibfield  {journal}
  {\bibinfo  {journal} {Eur. Phys. J. A}\ }\textbf {\bibinfo {volume} {25}},\
  \bibinfo {pages} {499} (\bibinfo {year} {2005})}\BibitemShut {NoStop}%
\bibitem [{\citenamefont {Honma}\ \emph {et~al.}(2009)\citenamefont {Honma},
  \citenamefont {Otsuka}, \citenamefont {Mizusaki},\ and\ \citenamefont
  {Hjorth-Jensen}}]{honma09}%
  \BibitemOpen
  \bibfield  {author} {\bibinfo {author} {\bibfnamefont {M.}~\bibnamefont
  {Honma}}, \bibinfo {author} {\bibfnamefont {T.}~\bibnamefont {Otsuka}},
  \bibinfo {author} {\bibfnamefont {T.}~\bibnamefont {Mizusaki}}, \ and\
  \bibinfo {author} {\bibfnamefont {M.}~\bibnamefont {Hjorth-Jensen}},\ }\href
  {\doibase 10.1103/PhysRevC.80.064323} {\bibfield  {journal} {\bibinfo
  {journal} {Phys. Rev. C}\ }\textbf {\bibinfo {volume} {80}},\ \bibinfo
  {pages} {064323} (\bibinfo {year} {2009})}\BibitemShut {NoStop}%
\bibitem [{\citenamefont {Petrovici}\ \emph {et~al.}(2018)\citenamefont
  {Petrovici}, \citenamefont {Andrei},\ and\ \citenamefont
  {Chilug}}]{petrovici18b}%
  \BibitemOpen
  \bibfield  {author} {\bibinfo {author} {\bibfnamefont {A.}~\bibnamefont
  {Petrovici}}, \bibinfo {author} {\bibfnamefont {O.}~\bibnamefont {Andrei}}, \
  and\ \bibinfo {author} {\bibfnamefont {A.}~\bibnamefont {Chilug}},\ }\href
  {\doibase 10.1088/1402-4896/aadec5} {\bibfield  {journal} {\bibinfo
  {journal} {Phys. Scr.}\ }\textbf {\bibinfo {volume} {93}},\ \bibinfo {pages}
  {114001} (\bibinfo {year} {2018})}\BibitemShut {NoStop}%
\bibitem [{\citenamefont {Honma}\ \emph {et~al.}(2004)\citenamefont {Honma},
  \citenamefont {Otsuka}, \citenamefont {Brown},\ and\ \citenamefont
  {Mizusaki}}]{honma04}%
  \BibitemOpen
  \bibfield  {author} {\bibinfo {author} {\bibfnamefont {M.}~\bibnamefont
  {Honma}}, \bibinfo {author} {\bibfnamefont {T.}~\bibnamefont {Otsuka}},
  \bibinfo {author} {\bibfnamefont {B.~A.}\ \bibnamefont {Brown}}, \ and\
  \bibinfo {author} {\bibfnamefont {T.}~\bibnamefont {Mizusaki}},\ }\href
  {\doibase 10.1103/PhysRevC.69.034335} {\bibfield  {journal} {\bibinfo
  {journal} {Phys. Rev. C}\ }\textbf {\bibinfo {volume} {69}},\ \bibinfo
  {pages} {034335} (\bibinfo {year} {2004})}\BibitemShut {NoStop}%
\bibitem [{\citenamefont {Bentley}\ \emph {et~al.}(2015)\citenamefont
  {Bentley}, \citenamefont {Lenzi}, \citenamefont {Simpson},\ and\
  \citenamefont {Diget}}]{bentley15}%
  \BibitemOpen
  \bibfield  {author} {\bibinfo {author} {\bibfnamefont {M.~A.}\ \bibnamefont
  {Bentley}}, \bibinfo {author} {\bibfnamefont {S.~M.}\ \bibnamefont {Lenzi}},
  \bibinfo {author} {\bibfnamefont {S.~A.}\ \bibnamefont {Simpson}}, \ and\
  \bibinfo {author} {\bibfnamefont {C.~A.}\ \bibnamefont {Diget}},\ }\href
  {\doibase 10.1103/PhysRevC.92.024310} {\bibfield  {journal} {\bibinfo
  {journal} {Phys. Rev. C}\ }\textbf {\bibinfo {volume} {92}},\ \bibinfo
  {pages} {024310} (\bibinfo {year} {2015})}\BibitemShut {NoStop}%
\bibitem [{\citenamefont {Petrovici}(2017)}]{petrovici17}%
  \BibitemOpen
  \bibfield  {author} {\bibinfo {author} {\bibfnamefont {A.}~\bibnamefont
  {Petrovici}},\ }\href {\doibase 10.1088/1402-4896/aa6d2f} {\bibfield
  {journal} {\bibinfo  {journal} {Phys. Scr.}\ }\textbf {\bibinfo {volume}
  {92}},\ \bibinfo {pages} {064003} (\bibinfo {year} {2017})}\BibitemShut
  {NoStop}%
\bibitem [{\citenamefont {Petrovici}(2018)}]{petrovici18}%
  \BibitemOpen
  \bibfield  {author} {\bibinfo {author} {\bibfnamefont {A.}~\bibnamefont
  {Petrovici}},\ }\href {\doibase 10.1103/PhysRevC.97.024313} {\bibfield
  {journal} {\bibinfo  {journal} {Phys. Rev. C}\ }\textbf {\bibinfo {volume}
  {97}},\ \bibinfo {pages} {024313} (\bibinfo {year} {2018})}\BibitemShut
  {NoStop}%
\end{thebibliography}%

\end{document}